\documentclass[aps,prd,onecolumn,groupedaddress,nofootinbib,preprint,11pt,a4paper]{revtex4-1}
\usepackage[colorlinks=true,citecolor=magenta,linkcolor=blue,breaklinks=true]{hyperref}
\usepackage{amsmath}
\usepackage{graphicx,graphics}
\usepackage{amssymb}
\usepackage{xcolor}
\usepackage{caption}
\usepackage[caption=false,labelformat=simple]{subfig}

\usepackage{booktabs}
\usepackage{cellspace}
\newcommand{\gsim}{\gtrsim}

\usepackage{feynmp-auto}
\usepackage{array}
\usepackage{cellspace}
\usepackage{multirow}
\begin{document}
\title{Probing the low mass pseudoscalar in flipped Two Higgs Doublet Model}


\author{Dilip Kumar Ghosh}
\email{tpdkg@iacs.res.in}
\affiliation{School of Physical Sciences, \\ Indian Association for the Cultivation of Science, \\ 2A \protect\& 2B, Raja S.C. Mullick Road, Kolkata 700032, India.}

\author{Biswarup Mukhopadhyaya}
\email{biswarup@iiserkol.ac.in}

\author{Sirshendu Samanta}
\email{ss21rs027@iiserkol.ac.in}

\author{Ritesh K. Singh}
\email{ritesh.singh@iiserkol.ac.in}
\affiliation{Department of Physical Sciences, \\ Indian Institute of
Science Education and Research Kolkata, \\ Mohanpur, 741246, India.}

\begin{abstract}
The phenomenology of the flipped two-Higgs-doublet model (2HDM) is relatively
less explored so far, as compared to the other, commonly discussed, types. It is
found that this scenario, like several others,  admits of a light neutral 
pseudoscalar $A$ in the mass range 20 - 60 GeV, consistently with all current experimental
data and theoretical constraints. However, the fact that such a pseudoscalar 
decays overwhelmingly into a $b\bar{b}$ pair makes its identification at the
Large Hadron Collider (LHC) a challenging task. After identifying the region
of the flipped 2HDM parameter space yielding a light pseudoscalar, we identify
a useful search channel in the process
 $pp \rightarrow A Z(Z^{*}) \rightarrow b\bar{b} \ell^+ \ell^-$. A cut-based analysis,
followed by one based on Boosted Decision Trees, shows that the light-$A$ scenario in flipped 2HDM should be detectable with rather high statistical
significance at the high-luminosity LHC run, even after including systematic
uncertainties. Furthermore, part of the parameter space, especially
around $m_A = 30 - 40$ GeV, is amenable to detection at the discovery level
within Run-2 itself.
\end{abstract}

\maketitle
\section{Introduction}
In the standard model (SM) of weak and electromagnetic interactions,
spontaneous breaking of the $SU(2)_L \times U(1)_Y$ local electroweak gauge symmetry into $U(1)_{EM}$
plays a central role, where {\it a single complex $SU(2)$ scalar doublet}
plays a crucial role\,\cite{Novaes:1999yn}. However, despite the discovery of a neutral spinless
particle with mass around 125 GeV broadly matching the Glashow-Salam-Weinberg
 Description\,\cite{Bilenky:1982ms}, the question as to whether it is `the Higgs' or `a Higgs' is yet to be
settled. Physicists have thus explored scenarios with more than one Higgs doublet (and
even larger $SU(2)$ multiplets)\,\cite{Branco:2011iw,Bhattacharyya:2015nca,Aggleton:2016tdd,Davoudiasl:2022mav,Ivanov:2017dad,Robens:2021lov}, perhaps encouraged by the observed replication of fermion families. The first step in such theorisation is two Higgs doublet models (2HDM)\,\cite{Branco:2011iw,Bhattacharyya:2015nca}.

The four major forms of 2HDM, aimed at suppressing flavour changing 
neutral currents (FCNC)\,\cite{Sher:2022aaa}, are classified as type I, type II, lepton-specific, and flipped 2HDM\,\cite{Branco:2011iw,Bhattacharyya:2015nca}.
They differ from each other in the formulation of the Yukawa interactions, with various
kinds of $\mathbb{Z}_2$ couplings to ensure that  both of the scalar doublets do not couple to both
$T_3 = +1/2$ and $-1/2$ fermion-antifermion pairs. Accelerator phenomenology of these different types
has been studied at various levels of detail, excepting the fourth category, namely, flipped 2HDM, on 
which, to the best of our knowledge,  only a limited number of studies have been performed\,\cite{Logan:2010ag,Hashemi:2018gfo,Dhargyal:2016eri}.
Here we report a study on the flipped 2HDM, with a specific theme.

The theme is woven around a light pseudoscalar physical state, which is present in all four of the aforementioned 
models. Such a particle cannot be very light in types I and II\,\cite{Mondal:2023wib,Atkinson:2021eox}, because of its unsuppressed
quark couplings. The lepton-specific scenario admits of a relatively light pseudoscalar
which dominantly couples to leptons, and has been studied in various contexts\,\cite{Chun:2018vsn,Ghosh:2020tfq,Dey:2021pyn,Mukhopadhyaya:2023akv,Mukhopadhyaya:2024nxm,Das:2024ekt}. This
work investigates the flipped 2HDM scenario, which, too, accommodates a light pseudoscalar (around 20 GeV
or more), satisfying all the direct search limits from the Large Hadron Collider (LHC). In view of the fact that 
suggestions about a relatively light spinless particle appear in the literature\,\cite{DJOUADI1991175,Chakdar:2021gpd}, such possibilities
are worthy of investigation in all models of extended scalar sectors.

The big challenge in unravelling the signature of a light pseudoscalar ($A$) in flipped 2HDM is that
it decays dominantly in the $b{\bar b}$ channel, which in general tends to get submerged in huge QCD
backgrounds. The solution lies in some associated production processes, which provide useful tags. Keeping this in mind, our investigation reveals that it is profitable to use the $AZ$ production channel, mediated by the lighter SM-like scalar $h$ (while the heavier scalar $H$, too, can mediate this, through its contribution is rather small 
in regions of the parameter space corresponding to a light $A$). The $Z$, via its leptonic decays, offers 
the much-needed tag.

The twist in the tale here is provided by the fact that even in this channel, a clear distinction of the $A$-signal
requires an invariant mass peak of the $b{\bar b}$-pair at $m_A$. This requires, among other things, a 
reasonably high $b$-jet identification efficiency, which in turn requires each $b$ to have hardness above a certain threshold. However, that
tends to suppress the event rates with an on-shell $Z$, once the acceptance cuts are imposed.
In other words, one is pushed to the edge of the allowed phase space of the AZ(Z$^*$) system. We find that the solution lies in sacrificing the demand for
an on-shell $Z$, demanding instead an invariant mass peak of $b{\bar b}$-pairs with adequate hardness, and looking for the signal \\
\vspace{-3.3em}
$$pp\rightarrow (b{\bar b})_{m_A} + \ell^+ \ell^-, ~~~(\ell = e, \mu) $$
We identify the regions corresponding to a light and detectable $A$ (20 - 50 GeV) in the flipped 
2HDM parameter space,
and find that a cut-based analysis itself can yield signals with statistical significance (including 10-20\%
systematics) range up to 15$\sigma$ at high-luminosity LHC (HL-LHC) even with an off-shell $Z$.
The significance improves further on carrying out an analysis based on Boosted Decision Trees (BDT)\,\cite{Roe:2004na}. 
The results of both kinds of analyses are reported here.

Two of the novel features of our analysis are as follows:
\begin{itemize}
    \item Existing studies in the low-mass region rely on fully hadronic $b\bar{b}$ decays, which suffer from overwhelming QCD backgrounds. In contrast, our analysis leverages the $Z^* \to \ell^+ \ell^-$ decay, offering a less background-prone final state and improved signal significance.
    \item The four final-state particles ($b\bar{b}\ell^+\ell^-$) reconstruct to a mass near 125 GeV, which serves as a natural discriminator against the dominant QCD background. This significantly enhances the signal-to-background ratio, making it an effective search strategy.
\end{itemize}


We stress upon the fact that the purpose of this investigation is to probe the case of light pseudoscalar and not a generic, exhaustive probe of the flipped 2HDM.
Because of this thematic focus,
we specifically identify regions where such a light pseudoscalar exists consistently with all phenomenological
and theoretical constraints. The Markov Chain Monte Carlo (MCMC) algorithm used to scan the parameter space has thus been subjected
to a bias in the low $m_A$ range, in addition to the aforesaid constraints. The relevant parameter regions
that emerge are therefore not covering the entire range of parameters in this model.
And we have selected some representative benchmark points out of these regions, which bring out the
viability of our proposed strategy.
  
The organization of the paper is as follows. A brief description of the flipped 2HDM and parameter space constraints is discussed in
Section \ref{sec:model}, serves as the justification for our chosen
benchmark points. Then, in section \ref{sec:collider}, we study the collider characteristics of the signal and backgrounds. We then reported results based on a cut-based approach and subsequently on Boosted Decision Trees(BDT), in sections \ref{sec:cutBased} and \ref{sec:BDT} respectively. We summarize and conclude in Section \ref{sec:summ}.

\section{flipped 2HDM: the model, its parameters and constraints} \label{sec:model}
Being a popular extension of the standard model (SM), the Two Higgs Doublet Model (2HDM) enriches the scalar sector with an additional complex $SU(2)_L$ doublet. Two scalar doublets are denoted by $\Phi_1$ and $\Phi_2$, and it gets the vacuum expectation value (vev) $v_1$ and $v_2$ respectively after Electro-weak-symmetry breaking (EWSB)\,\cite{Pich:2015lkh} with $v = \sqrt{v_1^2 + v_2^2}$, where $v$ is the {\it VEV} of the SM. Out of eight fields of the two complex scalar doublets, three fields are absorbed to give the mass of three gauge bosons ($Z, W^\pm$) and yield five physical scalar states with two CP-even neutral scalar ($h, H$), a CP-odd neutral scalar ($A$) and a pair of charged scalar ($H^\pm$). To avoid flavour changing neutral currents (FCNC)\,\cite{Sher:2022aaa} at the tree level, an extra $\mathbb{Z}_2$ symmetry is routinely imposed. This $\mathbb{Z}_2$ is defined in different ways, ensuring that no set of quarks or leptons, with $T_3$ = $\mathcal{+}$ or $\mathcal{-} \frac{1}{2}$, couples to both of the scalar doublets.

Neglecting CP-violation, the scalar potential can be written as:
\begin{eqnarray}
V_\text{scalar}
&=&  m_{11}^2 \Phi_1^\dagger \Phi_1
    + m_{22}^2 \Phi_2^\dagger \Phi_2
    + \lambda_1 \left(\Phi_1^\dagger \Phi_1\right)^2
    + \lambda_2 \left(\Phi_2^\dagger \Phi_2\right)^2 + \lambda_3 \left(\Phi_1^\dagger \Phi_1\right)
     \left(\Phi_2^\dagger \Phi_2\right) \nonumber \\
& & + \lambda_4 \left(\Phi_1^\dagger \Phi_2\right)
     \left(\Phi_2^\dagger \Phi_1\right) + \left\{ -m_{12}^2 \Phi_1^\dagger \Phi_2
    + \frac{\lambda_5}{2} \left(\Phi_1^\dagger \Phi_2\right)^2
    + \text{h.c.} \right\}, \label{eqn:LagS}
\end{eqnarray}
Although the quadratic term $m_{12}^2 \Phi_1^\dagger \Phi_2$ softly breaks the $\mathbb{Z}_2$ symmetry, it does not introduce the FCNC further. The mixing angle $\beta = \tan^{-1}(\frac{v_2}{v_1})$ arises during mass matrix diagonalization of charged scalar and pseudoscalar, whereas $\alpha$, the mixing angle for neutral scalar, can be expressed in terms of the model parameters.

Depending on the scalar-doublet fermions couplings in the Yukawa potential, the 2HDM can be categorized into four types (Type-I, Type-II, Type-X, and flipped). In this work, we are dealing with the flipped type 2HDM where, in the Yukawa interactions, down-type quarks couple with $\Phi_1$ and up-type quarks and leptons couple with $\Phi_2$. Thus, in the mass eigen basis, the Yukawa interactions can be written as\,\cite{Branco:2011iw}:
\begin{eqnarray}
      \mathcal{L}_\text{Yukawa}
&=& - \sum_{f} \frac{m_f}{v} \left(\xi_f^h \bar f f h
    + \xi_f^H \bar f f H - i \xi_f^A \bar f \gamma_5 f A \right)
      \nonumber\\
& & - \frac{\sqrt{2}}{v}\Big[V_{ud}^\text{CKM}
      \left(m_u\xi_u^A \bar{u}_R d_L
    + m_d\xi_d^A \bar{u}_L d_R\right)H^+
    + m_\ell \xi_\ell^A \bar{\nu}_L\ell_R H^+
    + \text{h.c.}\Big]\!,
\end{eqnarray}
where
\begin{eqnarray}
&& \xi_{u}^h = \frac{\cos\alpha}{\sin\beta}, \qquad
\xi_{u}^H = \frac{\sin\alpha}{\sin\beta}, \qquad
\xi_{u}^A = \cot\beta, \label{eqn:hqq-coup} \nonumber\\
&& \xi_{d}^h = -\frac{\sin\alpha}{\cos\beta}, \qquad
\xi_{d}^H = \frac{\cos\alpha}{\cos\beta}, \qquad
\xi_{d}^A = \tan\beta, \label{eqn:hqq-coup} \nonumber\\
&&\xi_\ell^h = \frac{\cos\alpha}{\sin\beta}, \qquad \xi_\ell^H = \frac{\sin\alpha}{\sin\beta}, \qquad \xi_\ell^A = -\cot\beta. \label{eqn:hll-coup}
\end{eqnarray}

The model has a total of ten real parameters: $\{m_{11}, m_{22}, m_{12}, \lambda_1, \lambda_2, \lambda_3, \lambda_4, \lambda_5, v_1, v_2\}$, of which six of them are independent, when we put $m_h \approx$ 125 GeV, and $\sqrt{v_1^2 + v_2^2} = \text{246 GeV}$ along with two tadpole equations of electroweak symmetry breaking 
conditions. For each value of the thus emerging independent parameters, one obtains $m_h$, $m_H, m_{H^\pm}, m_A$ as the masses of the light and the heavy neutral scalar, charged scalar, and pseudoscalar, respectively. The masses can now be expressed :
\begin{eqnarray}
      m_{H}^{2}~
&=&    M^2\,s_{\alpha-\beta}^2
    + \left(\lambda_1\,c^2_\alpha\,c^2_\beta
    + \lambda_2\,s^2_\alpha\,s^2_\beta\
    + \frac{\lambda_{345}}{2}\,s_{2\alpha}\,s_{2\beta}\right)v^2, \label{eqn:mH}\\
      m_h^2~\,
&=&   M^2\,c^2_{\alpha-\beta}
    + \left(\lambda_1\,s^2_\alpha\,c^2_\beta
    + \lambda_2\,c^2_\alpha\,s^2_\beta\
    - \frac{\lambda_{345}}{2} s_{2\alpha}\,s_{2\beta}\right)v^2,\\
      m_A^2~\,
&=&   M^2-\lambda_5\,v^2, \\
      m_{H^\pm}^2
&=&   M^2-\frac{\lambda_4+\lambda_5}{2} v^2 \label{eqn:mHm},
\end{eqnarray}
where $M^2 = m^2_{12}/(s_\beta\,c_\beta)$ and, for an angle $\theta$,
$s_\theta(c_\theta)$ represents $\sin\theta(\cos\theta)$, and $\lambda_{345} = \lambda_3 + \lambda_4 + \lambda_5$. When it comes to the phenomenology in the alignment limit,  $\sin ( \beta - \alpha ) \approx  1$\,\cite{Bernon:2015qea}.

The parameter space of flipped 2HDM, like all other extended EWSB sector scenarios, is constrained by
several theoretical issues and experimental results. We outline them below (for details, see\,\cite{Dey:2019lyr}).

\begin{description}
    \item[Vacuum stability] The demand for a stable minimum of the scalar potential at the EWSB scale requires\,\cite{Nie:1998yn} $\lambda_{1,2} > 0, \quad \lambda_3 > - \sqrt{\lambda_1\lambda_2}, \quad \text{and} \quad \lambda_3 + \lambda_4 -  |\lambda_5| > -\sqrt{\lambda_1\lambda_2}\,.$

    \item[Perturbative unitarity at the electroweak scale] It requires that the coupled-channel gauge and Higgs boson 2 $\rightarrow$ 2 scattering matrix should have its highest eigenvalue bounded above by the optical theorem\,\cite{PhysRevD.16.1519,PhysRevD.7.3111} for every partial wave. Also, $|\lambda_i| < 4\pi$ for every $i \in \{1,2,3,4,5ex\}$.

    \item[The data on the 125 GeV scalar] The signal strengths of the scalar in all channels should be constrained with LHC data at the 2$\sigma$-level. This is verified with the help of the { \tt HiggsSignals} package (which is a part of the {\tt HiggsTools})\,\cite{Bechtle:2020uwn,Bahl:2022igd}. For example, for low $m_A$, decays like $h \rightarrow A Z$ are constrained by { \tt HiggsSignals}.

    \item[Limits from search of new scalars] The limits coming from searches for the heavier neutral scalars in all channels are obtained from the code {\tt HiggsBounds} (also part of {\tt HiggsTools})\,\cite{Bechtle:2020pkv}.

    \item[Electroweak precision observables] These provide constraints via the limits described on the oblique electroweak parameters $S$, $T$, and $U$\,\cite{PhysRevD.46.381,ALEPH:2005ab,10.1093/ptep/ptaa104}.

    \item[Flavour observables] Contraints come mainly from $b \rightarrow s \gamma$\,\cite{HFLAV:2016hnz} and $B_s \rightarrow \mu^+ \mu^-$\,\cite{CMS:2014xfa}. These restrict mainly the $m_H^\pm$-$\tan \beta$ space.
\end{description}

\begin{figure}[h!]
    \centering    
    \subfloat[]{\includegraphics[width=0.5\textwidth]{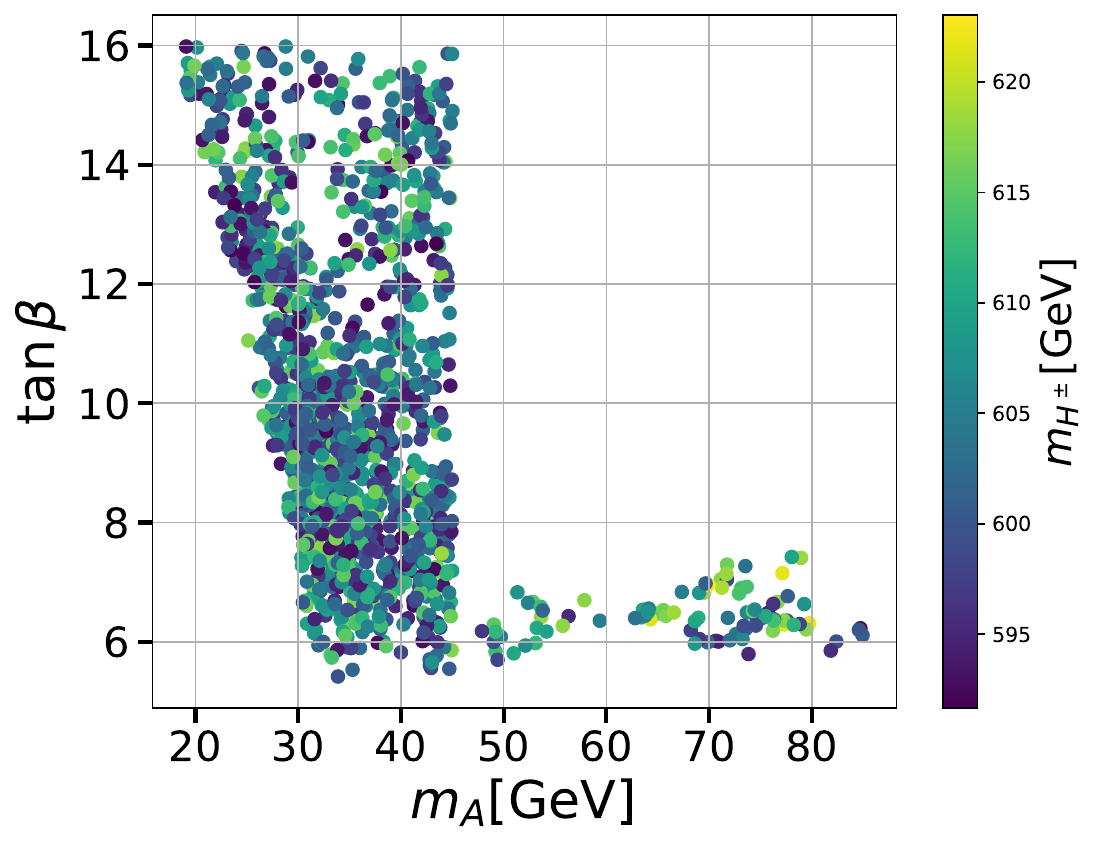}}\hfill
    \subfloat[]{\includegraphics[width=0.5\textwidth]{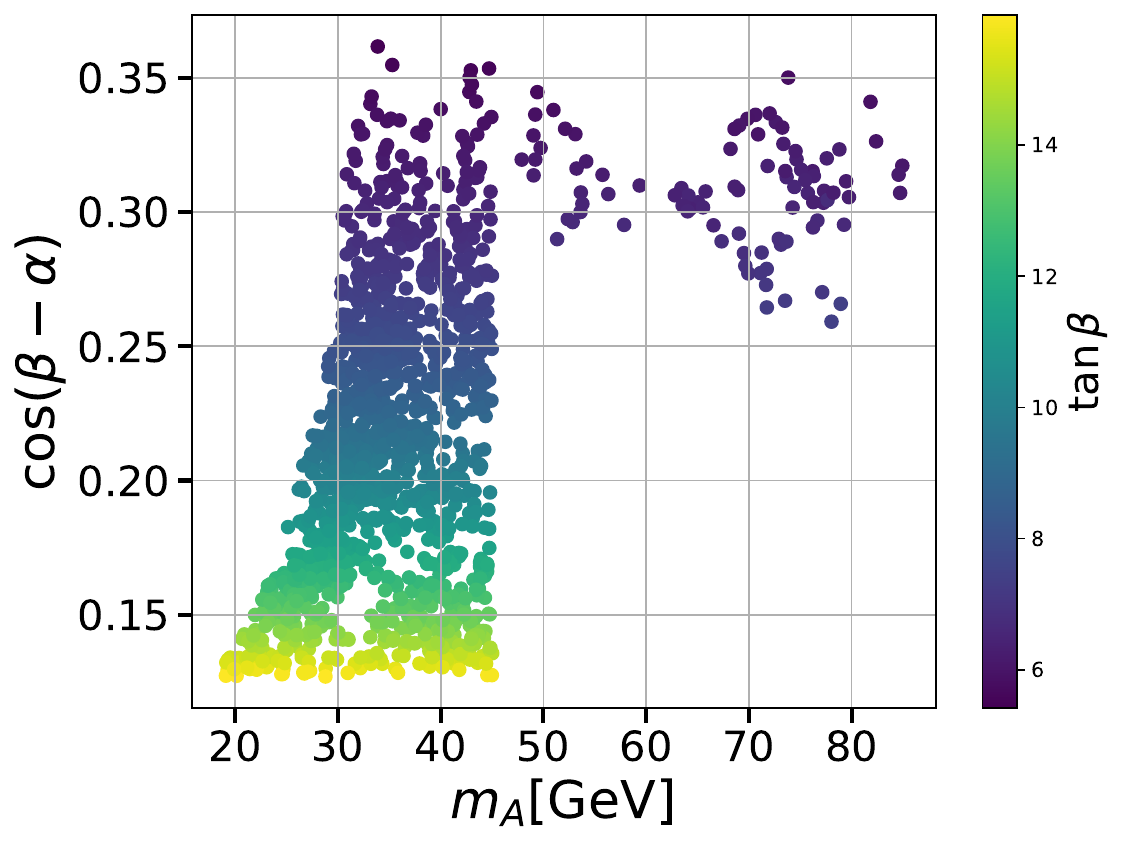}}
    \captionsetup{justification=raggedright,singlelinecheck=on}
    \caption{\it Allowed scattered points in $m_A$-$\tan \beta$ plane on the left panel with third axis $m_{H^\pm}$ (color-coded) and in $m_A$-$\cos(\beta-\alpha)$ plane on the right panel with third axis $\tan \beta$ (color-coded), once we impose all theoretical and experimental constraints discussed in the text.}
    \label{fig:AllowedParameter}
\end{figure}

We scan the parameter space of flipped 2HDM using a Markov Chain Monte Carlo(MCMC) algorithm, with a bias towards $m_A \in$ [15 GeV, 90 GeV], respecting all the constraints above. The MCMC scan finds the following ranges only for the allowed parameters:
$\lambda_1 \in [0.1,\ 1.6],\ 
\lambda_2 \in [0.05,\ 0.16],\ 
\lambda_3 \in [11.1,\ 12.5],\ 
\lambda_4 \in [-6.3,\ -4.3],\ 
\lambda_5 \in [5.6,\ 7.6],\ 
m_{12} \in [114,\ 285],\ 
\tan\beta \in [5.3,\ 26.6].$
The left panel of Figure~\ref{fig:AllowedParameter} shows a scatter plot of allowed $\tan \beta$ vs $m_A$ with $m_{H^\pm}$ along the color-coded third axis. On the right panel, the same points are plotted in $\cos(\beta-\alpha)$-$m_A$ plane with $\tan \beta$ along the color-coded third axis, where the remaining four parameters have been marginalised over. All the theoretical and experimental constraints discussed above have been used to filter the points that are included in the plot. As it is obvious from the Figure~\ref{fig:AllowedParameter}, $\tan\beta \lesssim$ 6-7 are consistent with all constraints for $ m_A\gtrsim$ 45 GeV. The benchmark points listed in Table~\ref{table:parameters} have been guided by this consideration, and they illustrate these regions in the parameter space, which makes an optimistic prediction for our suggested signal rate, as elaborated in the next section. 

It may be noticed that $\lambda_3$ is rather close to its perturbative limit at the
benchmark points used in our analysis. This is essentially due to the following
requirements: (a) $m_{H^\pm} \gtrsim 600$ GeV from flavour physics constraints,
(b) $|m_{H^\pm} -  m_{H}|$ must be small due to limits from precision electroweak
observables, (c) $m_h \approx 125$ GeV, and (d) $m_A \lesssim 50$ GeV in the context of our
investigation, where we are looking for {\em signals of a light pseudoscalar} in flipped 2HDM.
How the above constraints work can be understood from equations (\ref{eqn:mH}) - (\ref{eqn:mHm}). The apparent
alternatives, which can allow relatively small values of $\lambda_3$ are larger magnitudes of
$m_{12}$ and/or the quartic couplings $|\lambda_{4,5}|$. However, these options are limited by the
four aforementioned requirements. For example, $m_{H^\pm}$ or $m_{H}$ tend to get lowered below acceptable
limits, or $m_A$ ceases to be small compared to $m_{H^\pm}$. Note that relatively high values ($\gtrsim$6) of $\tan \beta$ have been chosen for $m_A$= 20 and 30 GeV. For higher values of $m_A$, substantial event rates are estimated only for $\tan \beta \lesssim 6$. The sharp line in Figure~\ref{fig:AllowedParameter} restricting higher $\tan \beta$ for $m_A \gtrsim 45$ is mainly due to limits from $g g \rightarrow A \rightarrow b \bar{b}$, which has been looked for recently in CMS\,\cite{CMS:2018pwl}.

We have already noted why one is pushed to a 'corner' of the parameter space where tree-level $\lambda_3$ is close to its' perturbative limit. This apparently brings in the possibility of the theory becoming non-perturbative around a TeV or so.  However, it is nonetheless interesting to explore the implications 
of a light pseudoscalar in flipped 2HDM, albeit from a purely phenomenological viewpoint, without
worrying about the UV completion of the theory. It may always be possible to have additional
degrees of freedom around the TeV scale, not yet accessible to the LHC, which temper the growth of the quartic couplings
as one reaches the TeV-scale threshold. Therefore, it is worthwhile to study the LHC implications
of this scenario, keeping in mind that any experimental evidence of a light pseudoscalar and
simultaneous hints of a flipped 2HDM may strongly suggest new degrees of freedom (possibly fermionic) around the TeV scale. Discussions on such possibilities are beyond the scope of the present study.

\begin{table}[h!]
\centering 
\begin{tabular}
{c@{\hspace{1cm}}c@{\hspace{1cm}}c@{\hspace{1cm}}c@{\hspace{1cm}}c@{\hspace{1cm}}c@{\hspace{1cm}}}
\hline
\toprule
 & \multicolumn{5}{c}{\textbf{Benchmark Points}} \\ [-10pt]
\textbf{Parameters} & & & \\ [-10pt]
 & \textbf{BP1} & \textbf{BP2} & \textbf{BP3} & \textbf{BP4} & \textbf{BP5}
 \\
\hline
\cmidrule(lr){2-4}
\midrule
$\lambda_1$ &   0.418  &   0.341  &  0.193  & 0.349 & 0.242  \\
$\lambda_2$ &  0.0645   &  0.0674   &  0.0676 & 0.0672 & 0.0664   \\
$\lambda_3$ &   11.50  &  12.40   &   11.70 & 11.80 & 12.10  \\
$\lambda_4$ &  -5.79  &  -5.29  &  -4.79 & -4.76 & -5.12  \\
$\lambda_5$ &   5.76  &   7.11  &   6.98 & 7.05 & 6.99 \\
$m_{12}$ (GeV)   &  152.0  &  256.0  &  266.0 & 272.0 & 259.0 \\
$\boldsymbol{\tan\beta}$ &  \textbf{15.0}  &  \textbf{6.4}  &  \textbf{5.8} & \textbf{5.6} & \textbf{6.2}  \\
\hline
$\boldsymbol{m_A}$ \textbf{(GeV)} &  \textbf{20.0}  &  \textbf{30.0}  &  \textbf{40.0} & \textbf{50.0} & \textbf{60.0}  \\
$\boldsymbol{\cos(\beta-\alpha)}$ &  \textbf{0.14}  &  \textbf{0.31}  &  \textbf{0.34} & \textbf{0.34} & \textbf{0.32}  \\
$\sin(\beta-\alpha)$ &  \textbf{0.99}  &  \textbf{0.95}  &  \textbf{0.94} & \textbf{0.94} & \textbf{0.95}  \\
$\mathcal{B}(h \rightarrow AZ)$ &  \textbf{13\%}  &  \textbf{14\%}  &  \textbf{0\%} & \textbf{0\%} & \textbf{0\%}  \\
$m_H$ (GeV) &  591.0  &  659.0  &  653.0 & 658.0 & 655.0  \\
$m_H^\pm$ (GeV) &  592.0  &  614.0  &  598.0 & 600.0 & 609.0  \\
\hline
\end{tabular}
\captionsetup{justification=raggedright,singlelinecheck=on}
\caption{\it The chosen parameter points for different benchmarks where the pseudoscalar mass is between 20 GeV and 60 GeV. The bold-faced parameters $\tan\beta$ and $m_A$ are the most important model parameters that decide the signal cross-section.}
\label{table:parameters}
\end{table}

\section{Collider Study} \label{sec:collider}
\subsection{Parton level characteristics of the Signal} 
\label{sec:signal}
\begin{figure}[h!]
    \centering
    \begin{fmffile}{higgs_triangular_loop}
    \begin{fmfgraph*}(180,83)
        \fmfleft{i1,i2}
        \fmfright{o1,o2,o3,o4}
        
        \fmflabel{$g$}{i1}
        \fmflabel{$g$}{i2}
        \fmflabel{$\ell^+$}{o1}
        \fmflabel{$\ell^-$}{o2}
        \fmflabel{$b$}{o3}
        \fmflabel{$\bar{b}$}{o4}
        
        \fmf{gluon}{i1,v1}
        \fmf{gluon}{i2,v2}
        
        \fmf{fermion,tension=0.5,left=0}{v1,v2}
        \fmf{fermion,tension=0.5,left=0,label=$t/b$}{v2,v3}
        \fmf{fermion,tension=0.5,left=0}{v3,v1}
        
        \fmf{dashes,label=$h_{\text{SM}}$}{v3,v4}
        
        \fmf{boson,label=$Z(Z^*)$,label.side=left,label.dist=0.4}{v4,v5}
        \fmf{dashes,label=$A$,label.side=right,label.dist=0.4}{v4,v6}
        
        \fmf{fermion}{v5,o1}
        \fmf{fermion}{o2,v5}
        
        \fmf{fermion}{v6,o3}
        \fmf{fermion}{o4,v6}
    \end{fmfgraph*}
    \end{fmffile}
    \captionsetup{justification=raggedright,singlelinecheck=on}
    \caption{\it Feynman diagram for the signal process where the SM Higgs is produced through Gluon-Gluon fusion (ggF) and decaying further into a $Z$ boson and a pseudoscalar.} 
    \label{fig:fyenmannDia}
\end{figure}
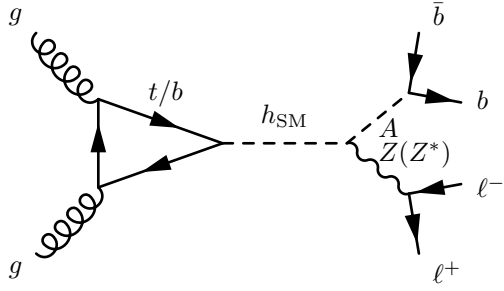

As discussed earlier, in the flipped 2HDM, there is no strong limit from the experimental side, forbidding a low mass pseudoscalar with $b\bar{b}$ as the dominant decay channel of the pseudoscalar with $\mathcal{B}$($A \rightarrow b \bar{b}) \approx 99\%$. However, such a decay channel faces a stiff challenge from backgrounds. Keeping this in mind, we consider the channel where the 125 GeV SM Higgs is produced via gluon fusion(ggf) before decaying into the $AZ(Z^*)$ mode, followed by the leptonic decay ($e, \mu$) of the real or virtual $Z$ boson. Figure~\ref{fig:fyenmannDia} represents the contributing Feynman diagram for the signal process. The presence of dileptons in the final states serves as a useful tag that keeps backgrounds under control. Since $(\beta-\alpha)\approx \pi/2$ in the alignment limit, the signal rates depend basically on $m_A$ and $\tan \beta$. The latter controls both the hAZ couplings ($\sim \cos (\beta-\alpha)$) and the branching ratio for $A \rightarrow b \bar{b}$ ($\sim \tan \beta$). It should be noted that the production diagram in Figure \ref{fig:fyenmannDia}, includes contributions from both $t$ and $b$-quarks. The $H$-mediated contribution is found not to measure up to the $h$-mediated one in spite of this. On the whole, the contribution of the heavier neutral scalar $H$ turns out to be at most 2\% of the
total contribution to our stipulated final state rate, although the $HZA$ coupling strength is
bigger than that for $hZA$ in the alignment limit. The primary reason for this is the kinematic suppression
of the $H$-mediated channel. Moreover, the $t\bar{t}h$ coupling dominates at the production level, despite the
inclusion of the b-quark triangle. Lastly, the $H$-induced contributions require higher $x$-values, so that
suppression of the parton distribution function takes place.\\

\begin{figure}[h!]
    \centering    
    \subfloat[]{\includegraphics[width=0.51\textwidth]{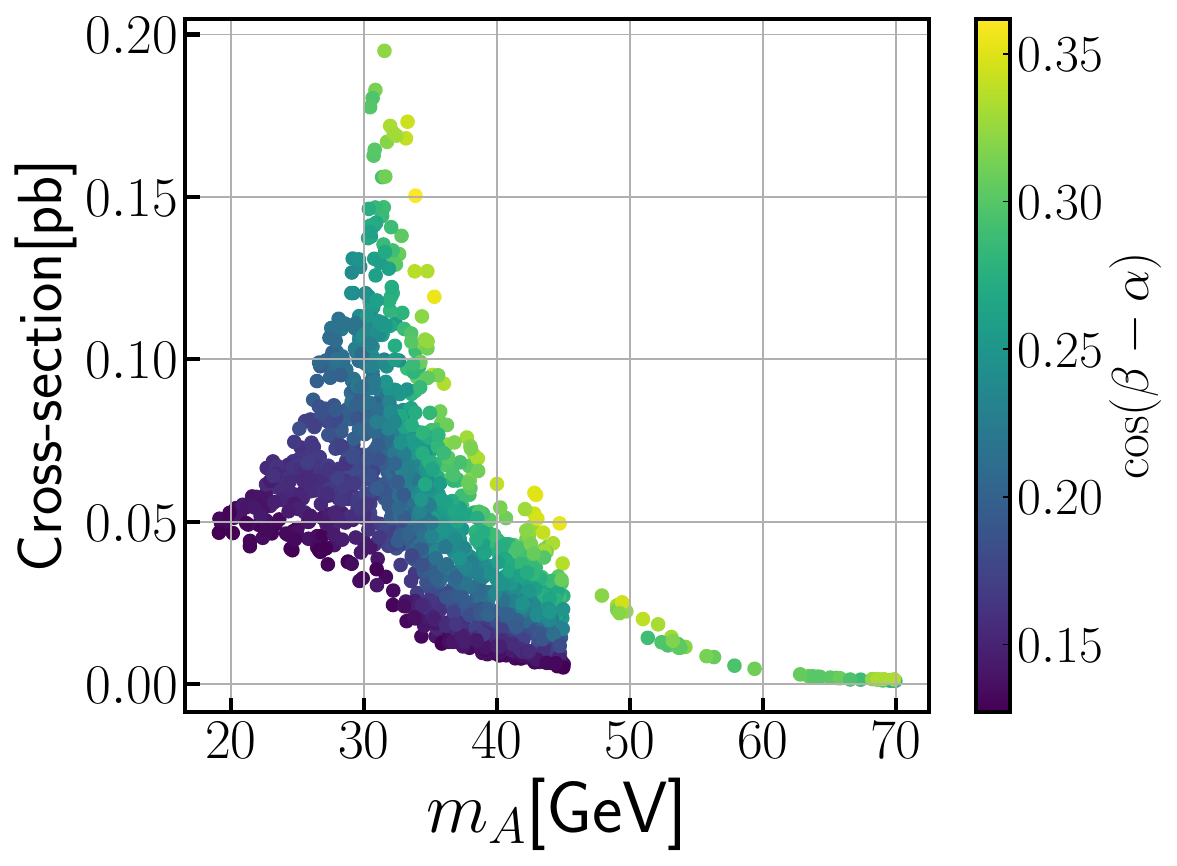}}\hfill
    \subfloat[]{\includegraphics[width=0.49\textwidth]{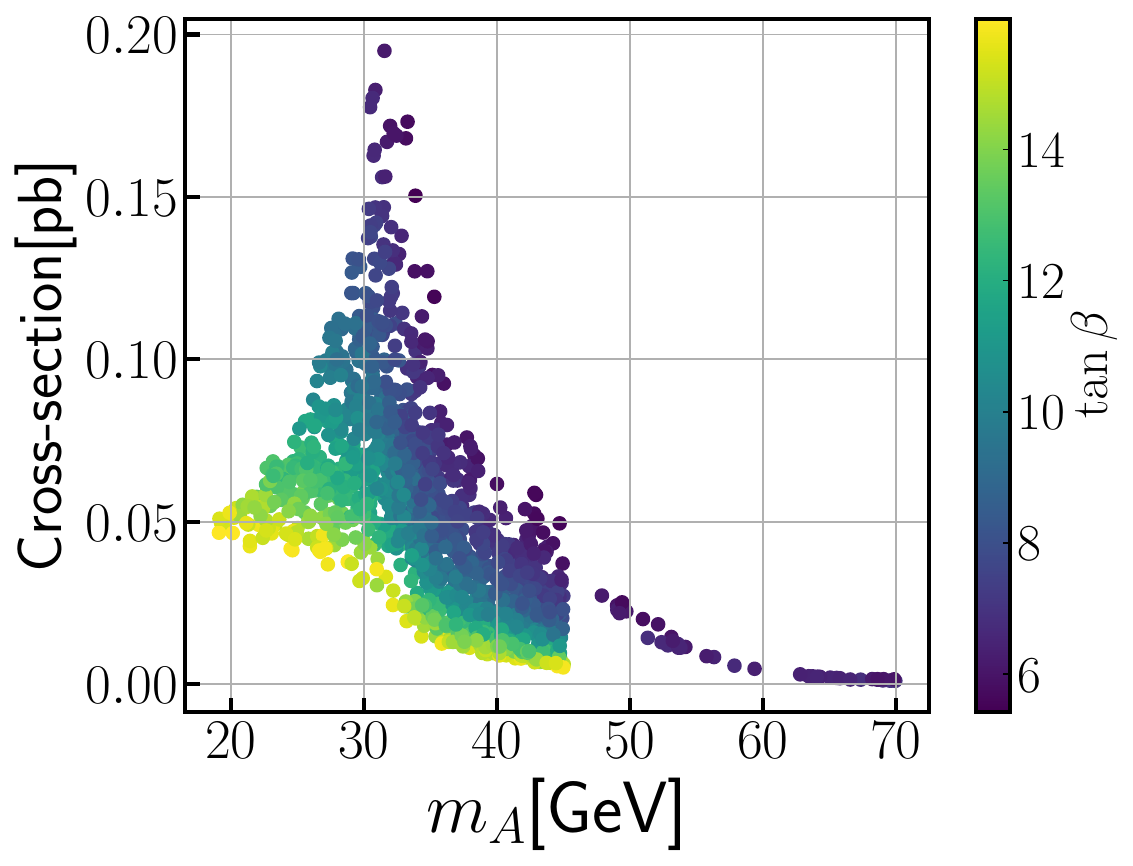}}
    \captionsetup{justification=raggedright,singlelinecheck=on}
    \caption{
    \it The left panel of the figure shows the variation of the cross-section (pb) for the process, $p p \rightarrow  AZ(Z^*)\rightarrow b\bar {b} \ell^+ \ell^-$ through the gluon-gluon fusion with third axis $\cos (\beta - \alpha)$ (color-coded). The right panel of the Figure represents the cross-section variation with the pseudoscalar mass keeping $\tan \beta$ on the third axis, which indicates the cross-section decreases with increasing $\tan \beta$ for all pseudoscalar masses.}
    \label{fig:parameterCross-sec}
\end{figure}

The left plot of the Figure~\ref{fig:parameterCross-sec} shows the cross-section for $pp\rightarrow AZ(Z^*) \rightarrow b\bar{b} \ell^+ \ell^-$ as a function of $m_A$, with $\cos(\beta -\alpha)$ as the third axis (color-coded) and the left plot with $\tan \beta$ as the third axis (color-coded). The plot corresponds to outputs from $\tt{Delphes}$\,\cite{deFavereau:2013fsa}, through which events generated at the parton level are passed after showering in $\tt{Pythia}$\,\cite{Bierlich:2022pfr}, imposing only the basic, trigger-level cuts. It shows that the rates fall sharply for $\tan \beta$ approaching 10 and beyond. Therefore, such a signal of a light pseudoscalar has hopes of distinction for low $\tan \beta$. This has influenced our choice of benchmark points. For bigger $m_A$ $(\approx 60)$ GeV, where there is a further kinematic suppression of the rate, the signal is detectable for the smallest possible values of $\tan \beta$ consistent with flavour constraints. \\
The right plot of Figure~\ref{fig:parameterCross-sec} demonstrates the likely detectability of the signal for different values of $m_A$, with $\tan \beta$ (color-coded) along the third axis. $m_A \approx$ 20 GeV yields somewhat low rates, because the final state particles often fail to satisfy the trigger. Similarly, kinematic suppression begins for $m_A \gtrsim $ 40 GeV. It is around $m_A \approx$ 30 GeV that the signal is of the highest strength\footnote{We shall see in the next section that, even in the ``unfavorable'' regions mentioned here, we obtain detectable signals with (a) the choice of appropriate cuts, and (b) the use of BDT.}.

\begin{figure}[h!]
    \centering    
    \includegraphics[width=0.353\textwidth]{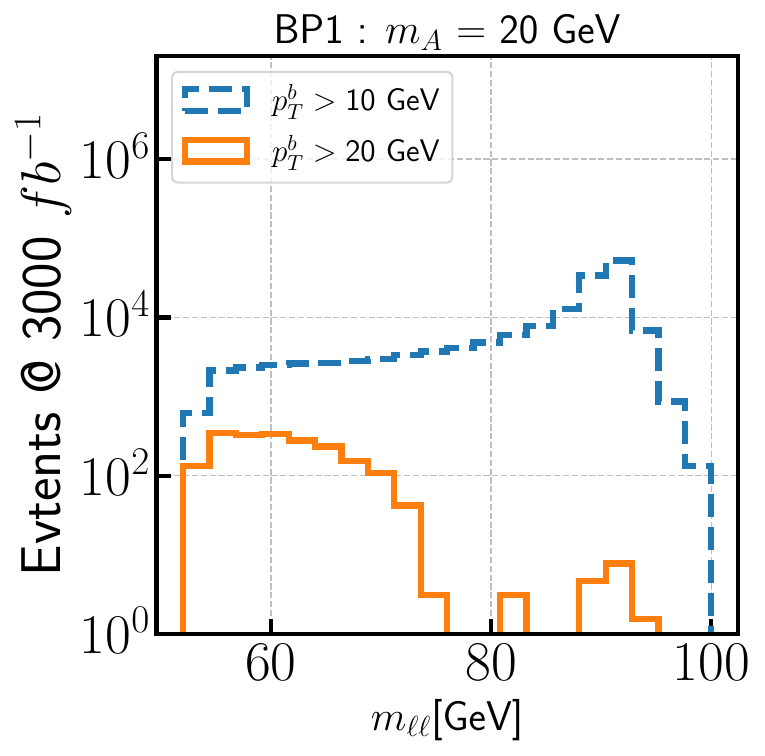}\hfill
    \includegraphics[width=0.323\textwidth]{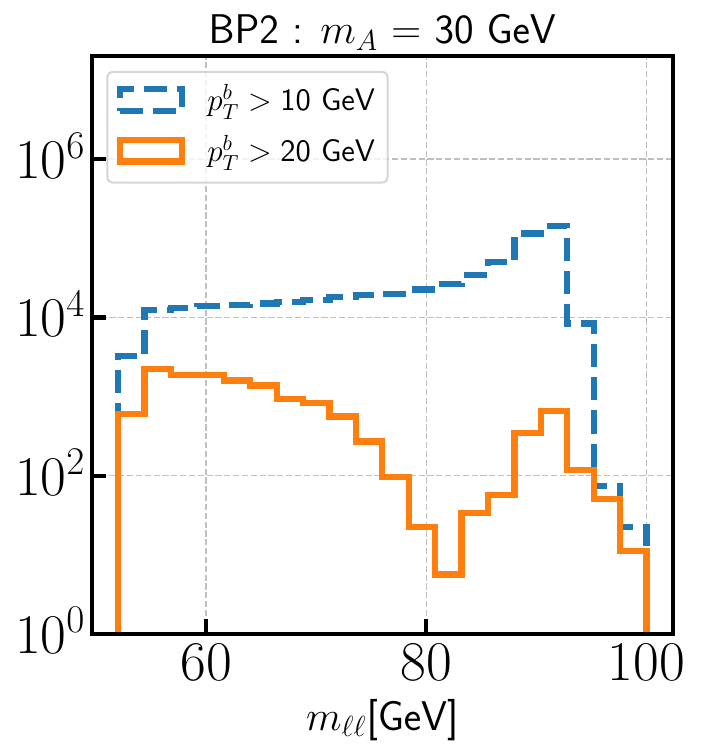}\hfill
    \includegraphics[width=0.323\textwidth]{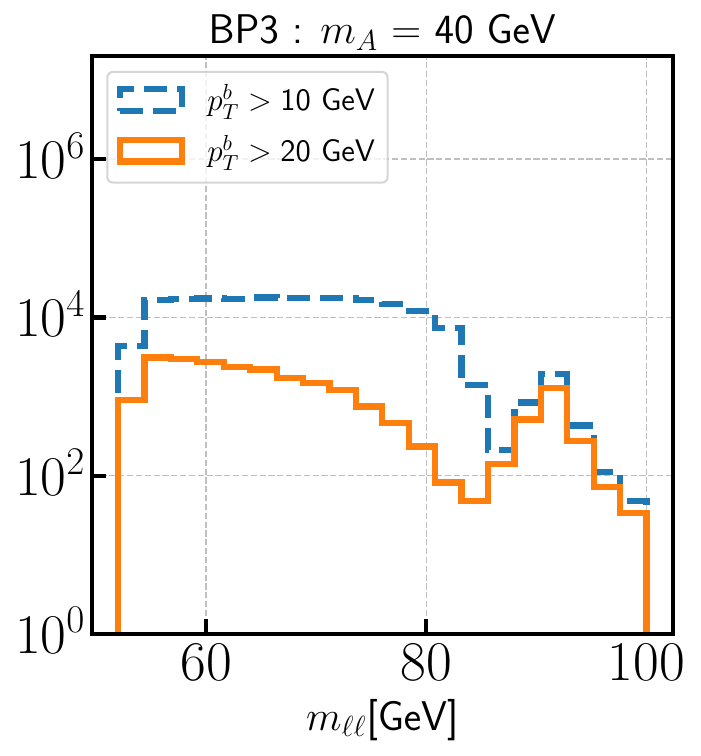}

    \captionsetup{justification=raggedright,singlelinecheck=on}
    \caption{\it Histogram of the invariant mass of leptons ($m_{\ell \ell}$) for three different signal benchmark points at the parton level. The Y-axis represents the number of events at 3000$fb^{-1}$ luminosity. The blue and orange histograms represent the number of events at the parton level with $p_T^b > \text{10 GeV}$ and $p_T^b > \text{20 GeV}$, respectively.}
    \label{fig:PartonOffShell}
\end{figure}

The relevance of some kinematic variables which prompt the choice of the acceptance cuts is indicated in Figure~\ref{fig:PartonOffShell}, where the lepton pair invariant mass distribution is shown at the parton level, for three different benchmark points. Each plot is generated using two distinct values of the transverse momentum $(p_T^b)$ trigger threshold for the $b$ quark.
We finally require that the total invariant mass of the two leptons and the two $b$-induced jets should exhibit a peak at $m_h$. This requires faithful reconstruction of the $b$-jets, for which a minimum $p_T^b$ of 20 GeV is more dependable than the initial parton-level trigger of 10 GeV. However, for final states resulting from an on-shell $h$, $p_T^b >$  20 GeV affects the hardness of each lepton. One thus notices a marked fall in the event rate for the $m_{\ell \ell}$ on the higher side, including the region corresponding to the $Z$-peak. The events for lower $m_{\ell \ell}$ are therefore preferable if one wants to retain more events with a $\ell \ell b\bar{b}$ peak that is traced us back to the 125 GeV scalar as the source. Therefore, we refrain from demanding an on-shell $Z$ as the source of the $\ell^+ \ell^-$ pair. The situation is different for $m_A \geq$ 30 GeV, for which there is scope for contributions from an on-shell $Z$. Here, the high A-mass not only relatively hard $b$'s, but also an on-shell $Z$ with $p_T$ on the higher side, which yields substantial contributions for $m_A$ = 40 GeV.

\subsection{Backgrounds} \label{sec:bkg}
\begin{figure}[h!]
    \centering
    \includegraphics[width=0.5\linewidth]{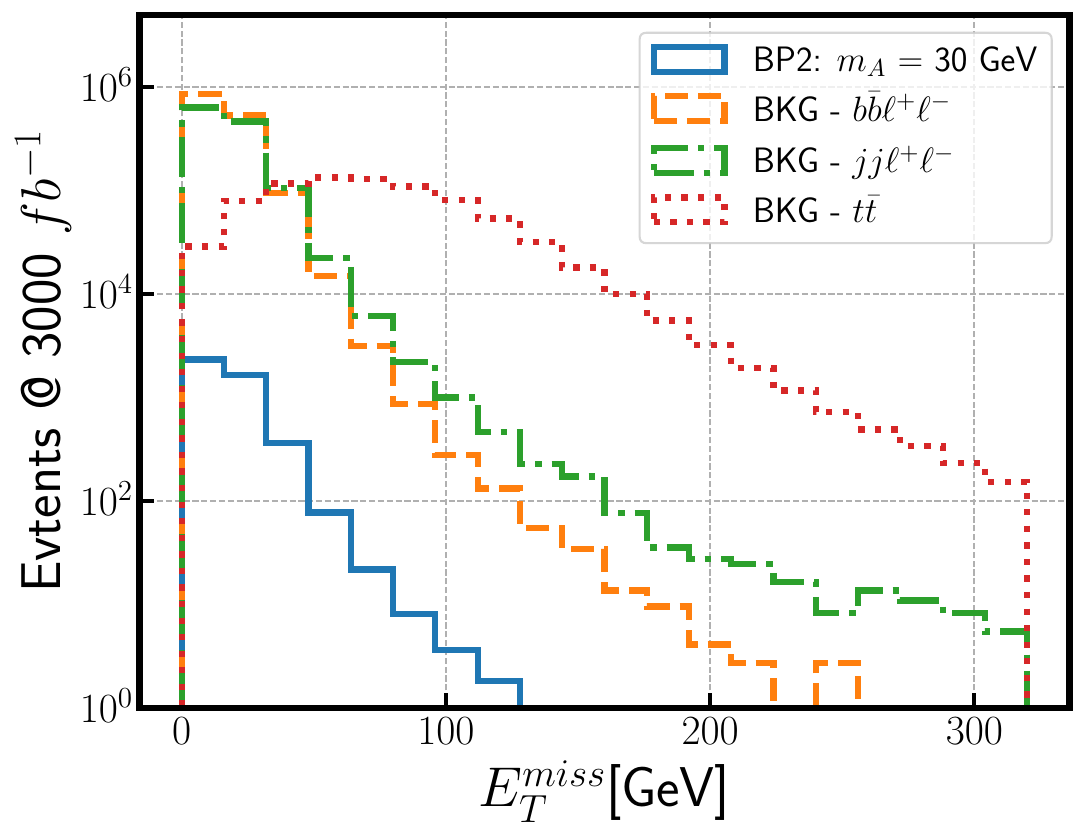}\hfill
    \includegraphics[width=0.5\linewidth]{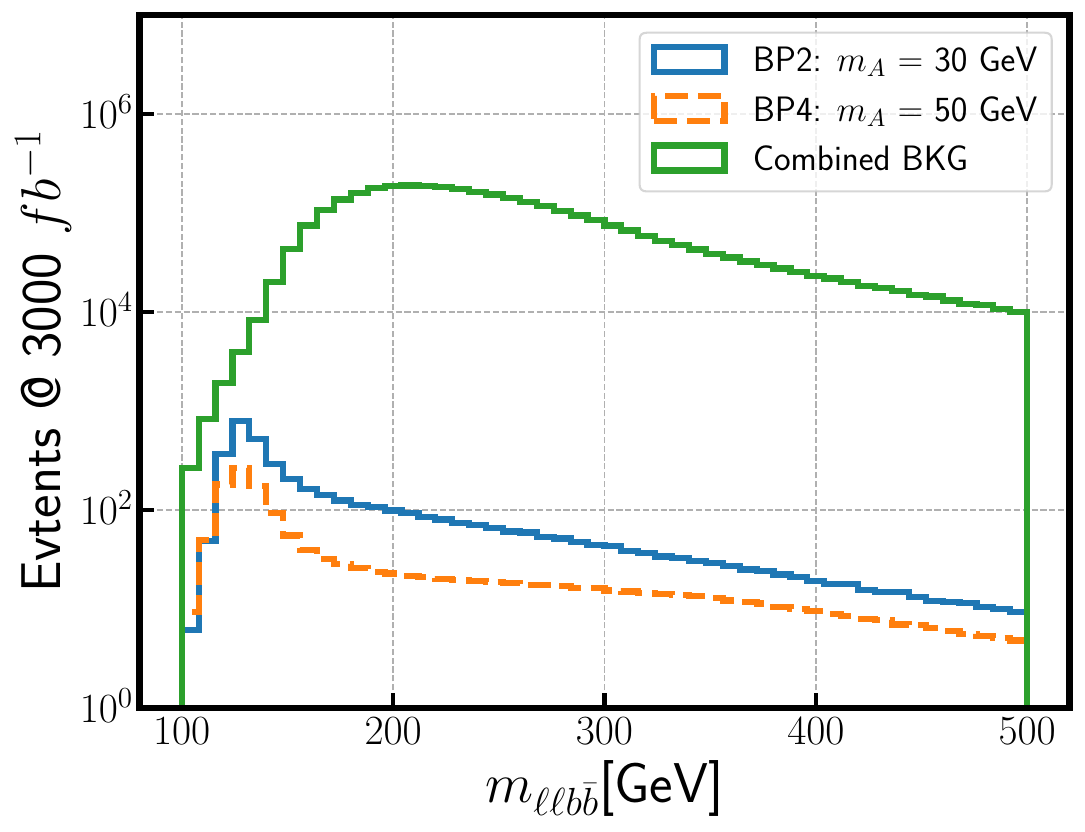}
    \captionsetup{justification=raggedright,singlelinecheck=on}
    \caption{\it The left panel shows the distribution of different backgrounds along with BP2, $m_A = \text{30 GeV}$. The signal lies in the lower side of the $E_T^{\rm miss}$; this variable kills the $t\bar{t}$ background to a large extent. Whereas the four-particle invariant mass is plotted in the right panel, where the signal peaks at the SM Higgs mass and the background deviates significantly and peaks near 210 GeV.}
    \label{fig:MET}
\end{figure}

While the signal rate has been computed at the leading order(LO), we have multiplied the LO estimates of the major backgrounds by the next-to-leading (NLO) $k$-factor. This makes our prediction of the signal conservative. The main sources of background we have considered are:
\begin{description}
    \item[$pp \to b\bar{b}\ell^+ \ell^-$] This background arises from the production of a vector boson ($V = Z, \gamma^*$) in association with a $b\bar{b}$ pair. From the production of the di-$Z$ boson, a pair of leptons and a pair of $b$-jets could appear. Similar particles can also appear from an off-shell photon associated with the $Z$-boson. A gluon splitting can also produce a pair of $b$-jets with the association of a vector boson. In a realistic detector, jets and leptons can be misidentified. So, the jet tagging efficiencies have been folded. These efficiencies depend on the jets' transverse momentum ($p_T$). In contrast, the lepton misidentification rate is negligible. Two $b$-jets are expected in the final state in the signal process, but misidentification can arise from multiple sources. The probability that a true $b$-jet is correctly tagged follows a $p_T$-dependent relation\,\cite{CMS:2012feb} which results in approximately 60\% tagging efficiency for the transverse momenta of $b$-jets in the range of 30–50 GeV, where the signal populates.

    \item[$pp \to jj\ell^+ \ell^-$] This reducible background involves light-flavor jets ($j = u, d, s, c, g$) produced in association with a vector boson. Though the jets are not $b$-jets, mis-tagging can cause this background to contaminate the signal region. Among light jets, charm jets are more likely to be mistakenly labeled as $b$-jets. Their mistagging rate is around 15\% for $p_T \approx 50~\text{GeV}$ follows a $p_T$-dependent relation\,\cite{CMS:2012feb}. For other light-flavor jets, the mistagging probability is significantly suppressed and approximately linear in $p_T$\,\cite{CMS:2012feb}, yielding a mistagging rate of about 1\% for jets with $p_T \approx 50$ GeV. Additionally, the probability of a light jet being misidentified as a lepton is highly suppressed, typically below 1\%.

    \item[$pp \to t\bar{t}$] Top quark pair production is another significant background due to its high cross section, in which the top quarks decay into bottom quarks and $W$ bosons, which decay into leptons of various flavors. However, semileptonic top quark decays result in missing transverse energy $(E_T^{\rm miss})$
    which offers a distinctive feature that suppresses this background.
    We apply a suitable selection that can significantly reduce this background because the signal process does not involve $E_T^{miss}$. The left panel of Figure~\ref{fig:MET} shows the $E_T^{miss}$ distribution for different processes in the background along with one benchmark point(BP2) of our signal.
\end{description}

Moreover, all these backgrounds can be further separated from the signal with an appropriate four-particle invariant mass ($m_{\ell \ell b\bar{b}}$) selection. The right plot of Figure~\ref{fig:MET} shows the distribution of $m_{\ell \ell b\bar{b}}$, indicating a clear separation of the peak between signal and backgrounds.

We generate the UFO model file for the event generation using the package {\tt SARAH}\,\cite{Staub:2013tta}. And use the {\tt SPheno}\,\cite{Porod:2011nf} package, which computes the Higgs masses, decay widths, branching ratios, and various couplings.  
At a center-of-mass energy of 14 TeV, we simulate all the signals and backgrounds at the parton level using the {\tt MadGraph5\_aMC@NLO}\,\cite{Alwall:2014hca}. Further, showering and hadronization are carried out using {\tt PYTHIA8}\,\cite{Bierlich:2022pfr} and {\tt Delphes}\,\cite{deFavereau:2013fsa} for detector simulation. The anti-$k_t$ technique\,\cite{Cacciari:2008gp} is used to cluster jets with a radius of 0.5. The NLO $k$-factors for the backgrounds $b \bar{b}\ell^+ \ell^-$, $jj\ell^+ \ell^-$ and $t \bar{t}$ are 1.7\,\cite{Frederix:2011qg}, 1.3\,\cite{Kim:2024ppt} 1.489\,\cite{Czakon:2013goa} respectively.

All the signals and backgrounds require triggers while passing through the detector simulator {\tt Delphes}. The following triggers in Equation~\ref{eqn:trigger} have been deployed to pass the events in the signals and backgrounds.
\begin{eqnarray}
{\tt Delphes} \text{ trigger:} \quad
   \left\{\ \begin{matrix}
   p_T^{\ell} > 10~\text{GeV},
&& |\eta_\ell| < 2.5, \\
   p_T^j > 20~\text{GeV},
&& |\eta_{j}| < 4.7, \\ &&\!\!\!\!\!\!\!\!\!\!\!\!\!\!\!\!\!\!\!\!\!\!\!\!\!\!\!\!\!\!\!\!\!\!\!\!\!\!\!\!\!\!\!\!\!\!\!\!\!\!\!\!\!\!\!\!\!\!\!\!\!\!\!\!\!\!\!\!\!\!\!\! \Delta R(j,j) > 0.4, \\
&&\!\!\!\!\!\!\!\!\!\!\!\!\!\!\!\!\!\!\!\!\!\!\!\!\!\!\!\!\!\!\!\!\!\!\!\!\!\!\!\!\!\!\!\!\!\!\!\!\!\!\!\!\!\!\!\!\!\!\!\!\!\!\!\!\!\!\!\!\!\!\!\! \Delta R(\ell,\ell) > 0.4, \\
&&\!\!\!\!\!\!\!\!\!\!\!\!\!\!\!\!\!\!\!\!\!\!\!\!\!\!\!\!\!\!\!\!\!\!\!\!\!\!\!\!\!\!\!\!\!\!\!\!\!\!\!\!\!\!\!\!\!\!\!\!\!\!\!\!\!\!\!\!\!\!\!\! \Delta R(j,\ell) > 0.4.

   \end{matrix} \right. \label{eqn:trigger}
\end{eqnarray}
Furthermore, the selection cut requires two opposite-sign same-flavour leptons and two $b$-jets. Table~\ref{tab:SelectionCutflow} shows the number of events for an integrated luminosity of 3000$fb^{-1}$ for all signals, together with the major backgrounds, after the {\tt Delphes} trigger and two different levels of the selection cut.

\begin{table}[!h]\renewcommand\arraystretch{1.25}
\begin{center}
\begin{tabular}{|c|c|c|c|c|c|c|c|c|}
\hline  & \multicolumn{8}{c|}{\textbf{Number of events at $\mathcal{L}=3000$~fb$^{-1}$ at $\sqrt{s}=14$~TeV LHC}}\\
\cline{2-9}
  & \multicolumn{5}{c|}{\textbf{Signals}} & \multicolumn{3}{c|}{\textbf{Backgrounds}} \\
\cline{2-9}
\textbf{Cuts} & \multicolumn{1}{c|}{\quad \textbf{BP1} \qquad}
& \multicolumn{1}{c|}{\quad \textbf{BP2} \qquad}
& \multicolumn{1}{c|}{\quad \textbf{BP3} \qquad}
& \multicolumn{1}{c|}{\quad \textbf{BP4} \qquad}
& \multicolumn{1}{c|}{\quad \textbf{BP5} \qquad}
& \multicolumn{1}{c|}{\quad $b \bar{b}\ell^+ \ell^-$ \qquad}
& \multicolumn{1}{c|}{\quad $jj\ell^+ \ell^-$ \qquad}
& \multicolumn{1}{c|}{$t \bar{t}$} \\
\cline{1-9}
~{\tt Delphes} \text{ trigger} (Equation~\ref{eqn:trigger})~~ & 0.15M  & 0.563M & 0.20M & 71.22K & 13.80K & 49.47M & 1450.0M & 13.31M\\
\hline
Selection Cut: $N_{\ell^\pm}$ = 2 & 75.12K & 0.26M & 85.30K & 28.33K & 5.64K & 25.0M &720.6M & 3.15M\\
\hline
Selection Cut: $N_{b-jets}$ = 2 & 978  & 4491 & 3000 & 1560 & 690 & 1.51M & 1.25M & 0.81M\\
\hline
\end{tabular}
\captionsetup{justification=raggedright,singlelinecheck=on}
\caption{\it Number of signal and different backgrounds events at the integrated luminosity 3000 $fb^{-1}$ passed after {\tt Delphes} trigger and two different selection cuts.}
\label{tab:SelectionCutflow}
\end{center}
\end{table}

\section{Results with Cut-Based Analysis} \label{sec:cutBased}
The signal channel has advantages in QCD background reduction because of two isolated leptons in the final state. 

\begin{figure}[h!]
    \centering
    \includegraphics[width=0.5\linewidth]{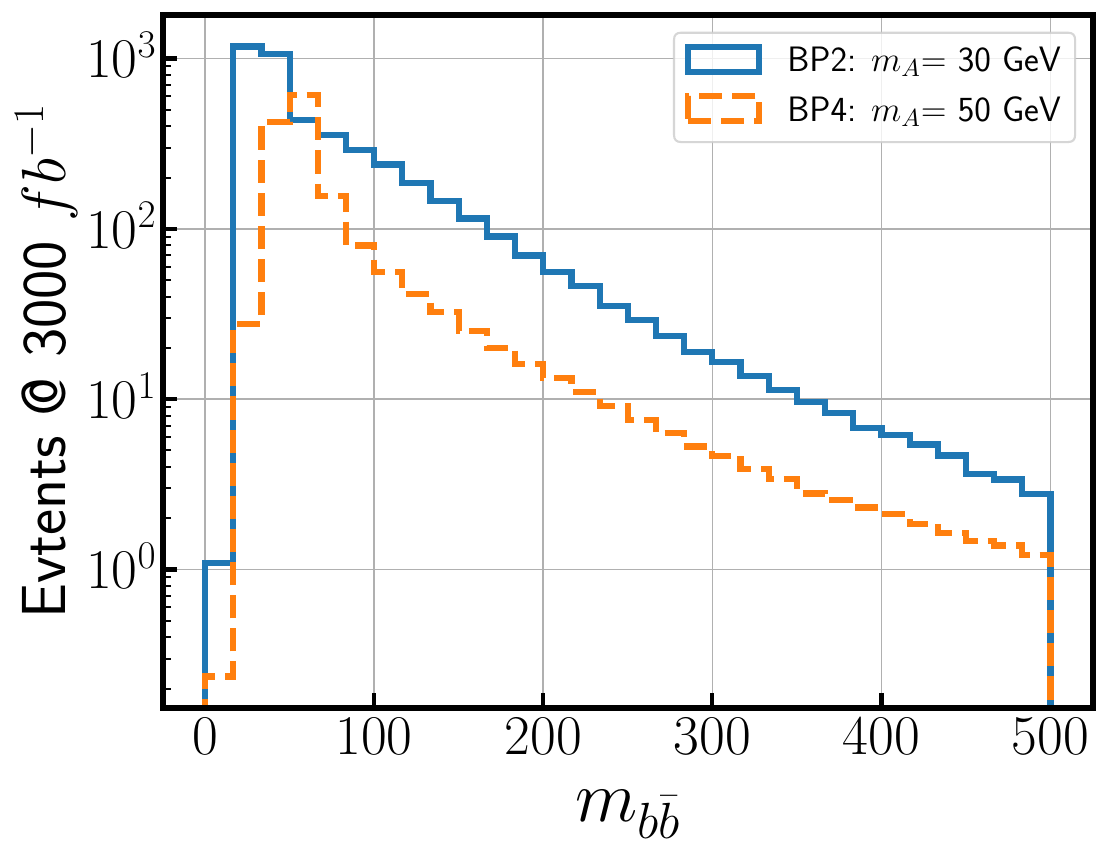}
    \captionsetup{justification=raggedright,singlelinecheck=on}
    \caption{\it The figure shows the distribution of the two b-jets invariant mass ($m_{b \bar{b}}$) for two different signal benchmark points(BP2, BP5).}
    \label{fig:m_bb}
\end{figure}

We perform our numerical analysis assuming an integrated luminosity of $\int \mathcal{L} dt = 3000~\text{fb}^{-1}$ in the context of the High-Luminosity LHC (HL-LHC) at a center-of-mass energy of 14 TeV. In Table~\ref{tab:cutflow}, we represent the number of signal events corresponding to our BPs that survive after each stage of cuts. For comparison, we also show the different number of background events in the same table. For distinguishing the signal from the background, one of the most effective kinematic variables is the invariant mass of the two $b$-jets and two leptons, denoted as $m_{\ell\ell b\bar{b}}$. The right side of Figure~\ref{fig:MET} shows the distribution of the $m_{\ell \ell b \bar{b}}$ for BP2, $m_A = \text{30~GeV}$, and BP4, $m_A = \text{50~GeV}$, along with all combined backgrounds. Where the signal peaks at the SM Higgs mass and the background population near 210~GeV. Although $m_{\ell\ell b\bar{b}}$ extends on the higher side with a long tail in signal, those events mainly come from mistagged $b$-jets or a $b$-jet produced after the parton shower whose mother is not the pseudoscalar. Those events that push the four-particle invariant mass higher are also accumulated in the long tail of the $m_{b\bar {b}}$ as shown in Figure~\ref{fig:m_bb}.

\begin{figure}
\subfloat[]{\includegraphics[width=0.48\textwidth]{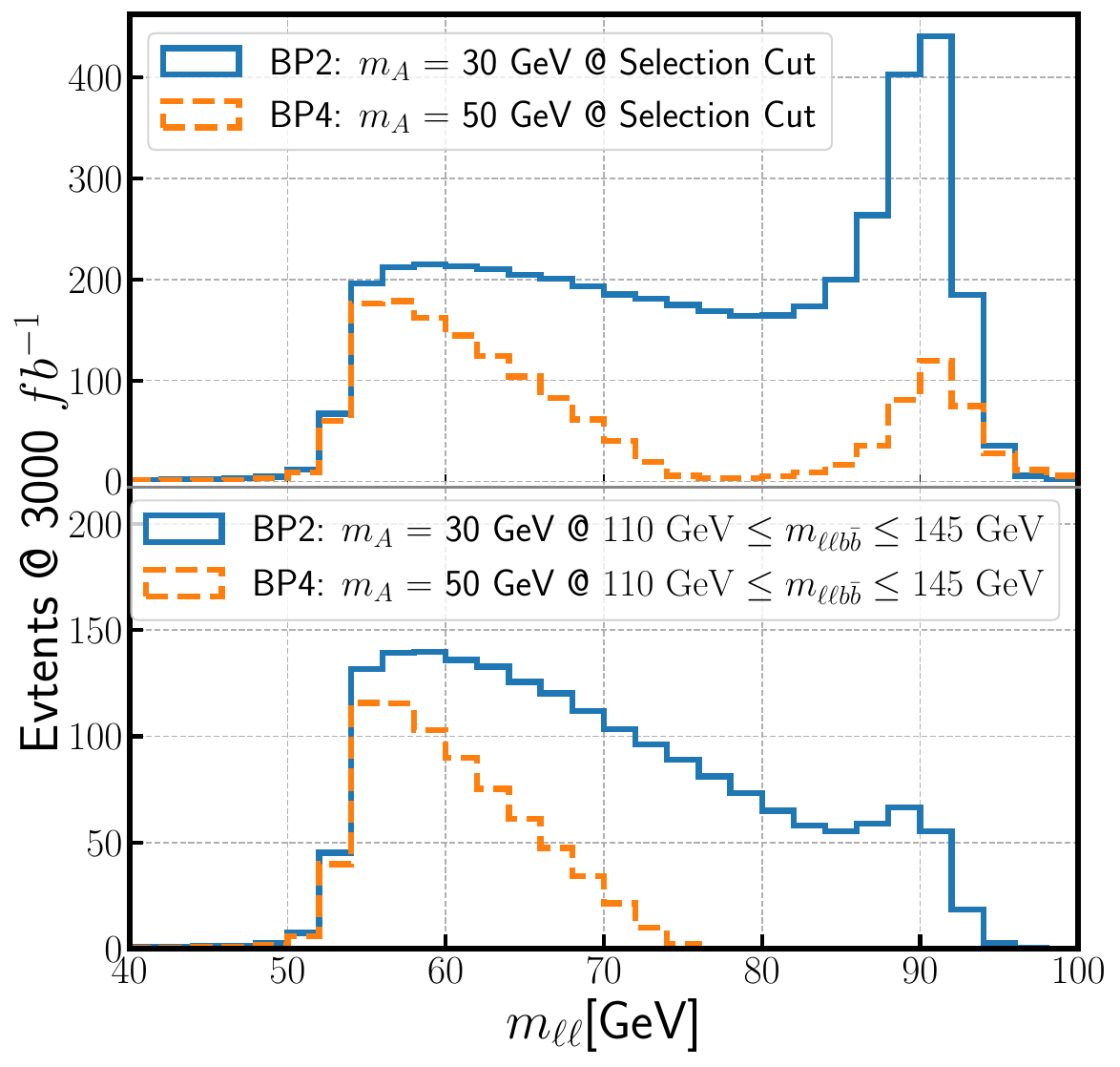}\label{fig:cutBased3}}\hfill
\subfloat[]{\includegraphics[width=0.48\textwidth]{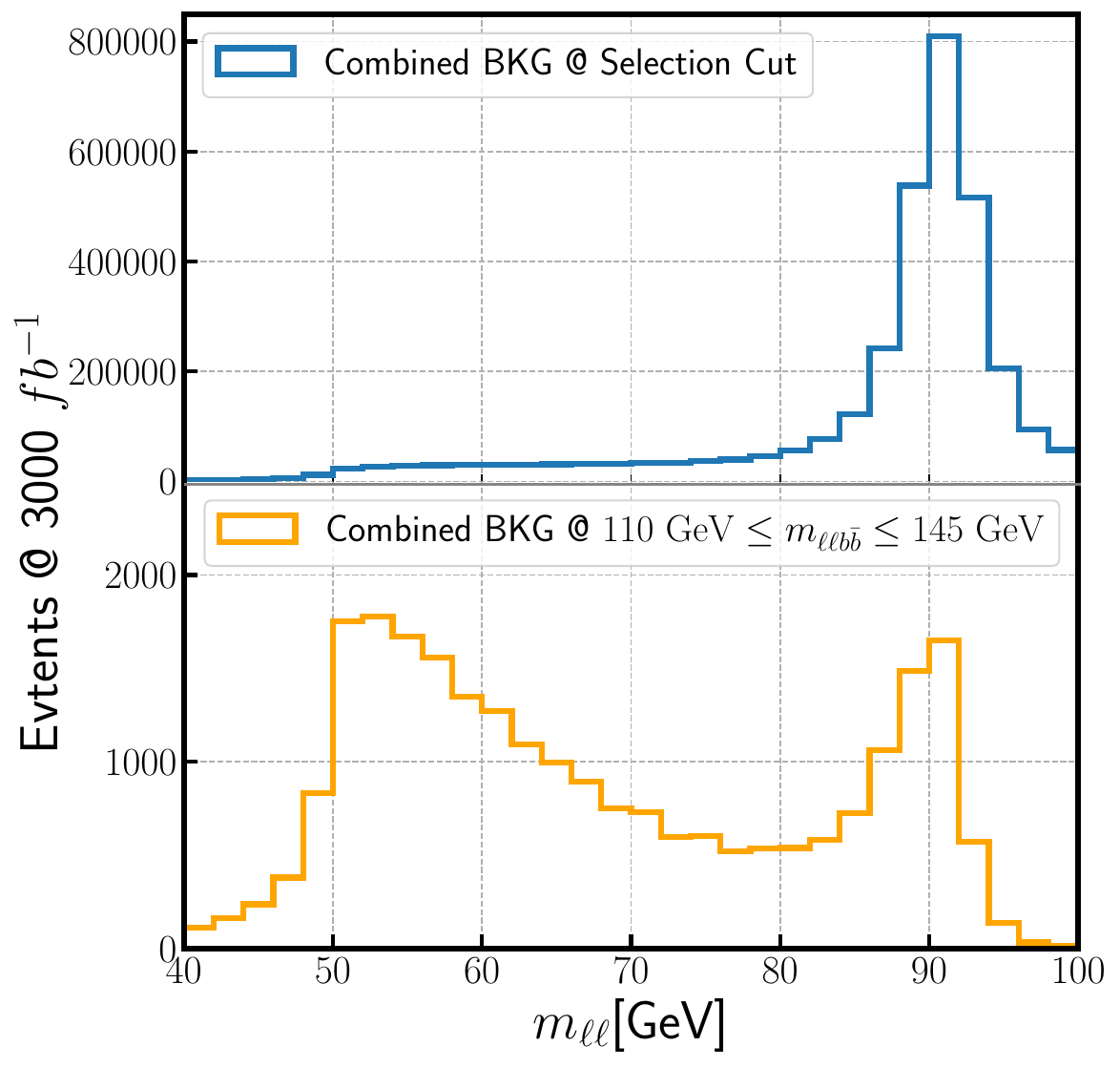}\label{fig:cutBased4}}
\captionsetup{justification=raggedright,singlelinecheck=on}
\caption{\it Two plots of the figure show the distribution of the $m_{\ell \ell}$ for two different signals (BP2, BP5) and background before and after imposing the $m_{\ell \ell b\bar {b}}$ cut. After imposing the $m_{\ell \ell b \bar{b}}$ cut around the SM Higgs mass, mainly the off-shell contribution of the $Z$ boson survives in the signal, which is consistent with the earlier discussion in section~\ref{sec:signal} using parton-level data.}
\label{fig:CutBasedPlot}
\end{figure}

Between 110 GeV and 145 GeV, we put the cut on $m_{\ell \ell b \bar{b}}$, and it turns out that the signal events achieve a high survival rate for all benchmark points, but only $ 0.70\%$ of events survive in the overall backgrounds. After applying this cut, the events in the signal corresponding to the on-shell $Z$ boson peak in the $m_{\ell\ell}$ distribution are significantly suppressed. The remaining events mainly originate from lepton pairs production via an off-shell $Z$ boson as shown in Figure~\ref{fig:cutBased3}. This behavior is consistent with the parton-level characteristics of the signal, as discussed in Section~\ref{sec:signal}, where high $p_T^b$ are typically associated with off-shell $Z$ boson. Figure~\ref{fig:cutBased4} represents the $m_{\ell \ell}$ distribution for the backgrounds before and after the $m_{\ell \ell b\bar {b}}$ cut. After imposing this cut on the background, the $Z$ peak also largely shrinks. Then we impose the missing energy cut, namely, $E_T^{miss} \leq \text{40 GeV}$, to eliminate the $t\bar{t}$ background, mainly as shown before in Figure~\ref{fig:MET}. After imposing this, the $ t\bar {t}$ background shrinks by more than 85\%, where signal events are barely suppressed. Lastly, as two $b$-jets in our signal are produced from the pseudoscalar, the invariant mass of these two $b$-jets helps further in improving the signal over backgrounds. We apply different $m_{b \bar{b}}$ cuts for different benchmark points, as shown in the cut-flow Table~\ref{tab:cutflow}.

\begin{table}[!h]\renewcommand\arraystretch{1.25}
\begin{center}
\begin{tabular}{|c|c|c|c|c|c|c|c|c|}
\hline  & \multicolumn{8}{c|}{\textbf{Number of events at $\mathcal{L}=3000$~fb$^{-1}$ at $\sqrt{s}=14$~TeV LHC}}\\
\cline{2-9}
  & \multicolumn{5}{c|}{\textbf{Signals}} & \multicolumn{3}{c|}{\textbf{Backgrounds}} \\
\cline{2-9}
\textbf{Cuts} & \multicolumn{1}{c|}{\quad \textbf{BP1} \qquad}
& \multicolumn{1}{c|}{\quad \textbf{BP2} \qquad}
& \multicolumn{1}{c|}{\quad \textbf{BP3} \qquad}
& \multicolumn{1}{c|}{\quad \textbf{BP4} \qquad}
& \multicolumn{1}{c|}{\quad \textbf{BP5} \qquad}
& \multicolumn{1}{c|}{\quad \textbf{$b \bar{b}\ell^+ \ell^-$} \qquad}
& \multicolumn{1}{c|}{\quad \textbf{$jj\ell^+ \ell^-$} \qquad}
& \multicolumn{1}{c|}{\textbf{$t \bar{t}$}} \\
\cline{1-9}
~{\tt Delphes} trigger + Selection Cuts~~ & 978  & 4491 & 3000 & 1560 & 690 & 1.51M & 1.25M & 0.81M\\
\hline
$110 \text{ GeV} \leq m_{\ell \ell b \bar{b}} \leq 145 \text{ GeV}$ & 314 & 1922  & 1510 & 716 & 181 & 9367 & 6573 & 9320\\
\hline
$E_T^{miss} \leq 40 \text{ GeV}$ & 305 & 1855 & 1457 & 697 & 177 & 9265 & 6408 & 1033\\
\hline
$10 \text{ GeV} \leq m_{b \bar{b}} \leq 30 \text{ GeV}$ & \textbf{267} & 583 & 45 & 7 & 1 & 1358 & 1910 & 134\\
$20 \text{ GeV} \leq m_{b \bar{b}} \leq 60 \text{ GeV}$ & 256 & \textbf{1834} & 1419 & 567 & 79 & 6804 & 5133 & 666\\
$25 \text{ GeV} \leq m_{b \bar{b}} \leq 65 \text{ GeV}$ & 123 & 1749 & \textbf{1431} & 646 & 120 & 7105 & 4663 & 724\\
$35 \text{ GeV} \leq m_{b \bar{b}} \leq 75 \text{ GeV}$ & 14 & 601 & 1272 & \textbf{680} & 166 & 6861 & 3627 & 747 \\
$45 \text{ GeV} \leq m_{b \bar{b}} \leq 85 \text{ GeV}$ & 7 & 89 & 527 & 591 & \textbf{174} & 5934 & 2600 & 652 \\
\hline
\end{tabular}

\captionsetup{justification=raggedright,singlelinecheck=on}
\caption{\it This is the cut-flow table for the different benchmark points. The $m_{\ell \ell b \bar{b}}$ and $E_T^{\text{miss}}$ cuts are applied similarly for each benchmark point, but the invariant mass of two b-jets ($m_{b \bar{b}}$) is chosen according to the pseudoscalar mass involved in the different benchmark points which are shown diagonally in bold. The number of events is specified for an integrated luminosity of 3000~fb$^{-1}$ for the corresponding cuts for signal and background processes.}
\label{tab:cutflow}
\end{center}
\end{table}

Various sources of uncertainty, such as luminosity uncertainty, parton distribution function (PDF) uncertainty, etc, can affect collider experiments. We anticipate some degree of uncertainty in the HL-LHC as well. We, therefore, choose to present the signal significance with systematic uncertainties :
\begin{eqnarray}
\mathcal{S} 
  =\sqrt{2} \left[ (s + b) \ln \left( \frac{(s + b)(b + \sigma_b^2)}{b^2 + (s + b) \sigma_b^2} \right) - \frac{b^2}{\sigma_b^2} \ln \left( 1 + \frac{\sigma_b^2 s}{b(b + \sigma_b^2)} \right) \right]^{\frac{1}{2}}\label{eqn:significance},
\end{eqnarray}
Where $s$ and $b$ are signal and background event numbers. The systematic uncertainty in the background $\sigma_b^2$ is proportional to $b$ and can be given as a percentage of the background. We tabulate the signal significance for all the benchmark points in Table~\ref{tab:signi}, considering 10\% systematics, with benchmark-specific $m_{b\bar{b}}$ cuts applied to all the points. Results with 20\% systematics uncertainty are also shown there, where, in addition, the effects of the $m_{b\bar{b}}$ cut for each BP on other BPs are also shown. The expected signal significance for all the benchmark points is quite good. 

\begin{table}[!h]\renewcommand\arraystretch{1.35}
\centering
\begin{tabular}{|c|>{\centering\arraybackslash}p{1.25cm}|
                   >{\centering\arraybackslash}p{1.25cm}|
                   >{\centering\arraybackslash}p{1.25cm}|
                   >{\centering\arraybackslash}p{1.25cm}|
                   >{\centering\arraybackslash}p{1.25cm}|}
\hline
& \multicolumn{5}{c|}{\textbf{Significance ($\mathcal{S}$) at 10\% systematics}} \\
\cline{2-6}
\quad \textbf{$m_{b \bar{b}}$ Cut Dependence} \qquad
& \multicolumn{1}{c|}{\textbf{BP1}}
& \multicolumn{1}{c|}{\textbf{BP2}}
& \multicolumn{1}{c|}{\textbf{BP3}}
& \multicolumn{1}{c|}{\textbf{BP4}}
& \multicolumn{1}{c|}{\textbf{BP5}} \\
\cline{1-6}
\textbf{BP's specific cut} & 4.3 & 15.1 & 11.9 & 6.0 & 1.7 \\
\hline
& \multicolumn{5}{|c|}{\textbf{Significance ($\mathcal{S}$) at 20\% systematics}} \\
\hline
$10 \text{ GeV} \leq m_{b \bar{b}} \leq 30 \text{ GeV}$ & \textbf{4.1} & 8.8 & 0.7 & 0.1 & 0.01\\
$20 \text{ GeV} \leq m_{b \bar{b}} \leq 60 \text{ GeV}$ & 2.0 & \textbf{14.5} & 11.2 & 4.5 & 0.6\\
$25 \text{ GeV} \leq m_{b \bar{b}} \leq 65 \text{ GeV}$ & 1.0 & 13.9 & \textbf{11.4} & 5.2 & 0.9\\
$35 \text{ GeV} \leq m_{b \bar{b}} \leq 75 \text{ GeV}$ & 0.1 & 5.1 & 10.7 & \textbf{5.6} & 1.4\\
$45 \text{ GeV} \leq m_{b \bar{b}} \leq 85 \text{ GeV}$ & 0.06 & 0.8 & 4.9 & 5.5 & \textbf{1.6}\\
\hline
\end{tabular}
\captionsetup{justification=raggedright,singlelinecheck=on}
\caption{\it This is the signal significances at the $\sqrt{s} = 14$~TeV LHC at
$\mathcal{L} = 3000$~fb$^{-1}$, for two different systematics 10\% and 20\%. For the $10\%$ case, a benchmark-specific $m_{b \bar{b}}$ cut is applied to determine the signal significance, while for $20\% $ case,
all signal significance has been calculated using different $m_{b \bar{b}}$ cuts for all different benchmark points.}
\label{tab:signi}
\end{table}
The signal significance is strongest at about 15$\sigma$ with 10\% systematic at 3000 fb$^{-1}$ luminosity for the BP2, where the pseudoscalar mass is 30 GeV. This signal can be seen at 330 fb$^{-1}$ luminosity with 10\% systematic at the discovery level 5$\sigma$. As the pseudoscalar mass increases, the signal relevance decreases. The signal significance for a 50 GeV pseudoscalar is near 6$\sigma$. For the pseudoscalar mass 60 GeV, it drops to 1.7$\sigma$. But, as will be covered in the next section, applying the Boosted Decision Tree (BDT) method would improve even more.

\section{Result Using BDT} \label{sec:BDT}
In the collider, distinguishing signal events from the backgrounds is the fundamental challenge if the signal rate is sufficiently low. In our work, we see earlier that the simple cut-based analysis is not effective for the BP5, $m_A=60~\text{GeV}$. Machine learning techniques, on the other hand, are powerful tools to enhance signal discrimination by capturing the complex correlation between multiple kinematic variables in the signal and backgrounds over different phase spaces where the cut-based approach mostly fails. In this work, we have used the Boosted Decision Tree (BDT) as a machine learning model. BDT works by iteratively training multiple decision trees on a set of variables and captures the nonlinear correlations.

An important step in the BDT analysis is the selection of the discriminating variables, which could help separate the signal from the background. The most relevant kinematic variables include:
\begin{itemize}
    \item Four-body Invariant mass ($m_{\ell \ell b\bar{b}}$): The reconstructed four-body invariant mass in all benchmark signals peaks near 125-GeV SM Higgs mass. So, in the four-body invariant mass, we allow only those events that lie between 100 GeV and 160 GeV before passing to the BDT. It helps BDT recognize the signal and background pattern better.
    
    \item Two-leptons Invariant mass ($m_{\ell \ell}$): As in our signal, two leptons produced mainly from an off-shell Z-boson would be a great discriminator for the BDT.
\end{itemize}

Additionally, we pass the many different features, such as two-body invariant masses $m_{ij}$, pairwise distances in the $\eta$-$\phi$ plane $\Delta R_{ij}$, azimuthal angle separation $\Delta \phi_{ij}$, for $i,j \in \{\ell_1, \ell_2, b_1, b_2\}$. We also add missing energy ($E_T^{miss}$) and the scalar sum of the $p_T$ of all visible particles ($H_T$). It is trained using a simulated dataset where events are classified as a signal or background.  We employ {\tt XGBoost}\,\cite{DBLP:journals/corr/ChenG16}, an optimized gradient-boosting algorithm, to construct the decision trees. The set of hyperparameters employed for training the XGBoost classifier is summarized in Table~\ref{tab:xgb_hyperparams}.

\begin{table}[htbp]
    \centering
    \vspace{0.3em}
    {
    \begin{tabular}{|l|c|l|c|}
        \hline
        \textbf{Hyperparameter} & \textbf{Value} & \textbf{Hyperparameter} & \textbf{Value} \\
        \hline
        \texttt{objective}         & \texttt{binary:logistic} & \texttt{n\_estimators}     & 1000 \\
        \texttt{eval\_metric}      & \texttt{logloss}         & \texttt{learning\_rate}    & 0.01 \\
        \texttt{max\_depth}        & 3                        & \texttt{min\_child\_weight} & 5 \\
        \texttt{gamma}             & 1                        & \texttt{subsample}         & 0.6 \\
        \texttt{colsample\_bytree} & 0.6                      & \texttt{reg\_alpha}        & 1.0 \\
        \texttt{reg\_lambda}       & 5.0                      & \texttt{tree\_method}      & \texttt{hist} \\
        \hline
    \end{tabular}
    \captionsetup{justification=raggedright,singlelinecheck=on}
    \caption{ \it Hyperparameters used in training the XGBoost classifier. These were selected based on iterative tuning to balance performance and overfitting.}
    \label{tab:xgb_hyperparams}
    }
\end{table}

\begin{figure}[h!]
    \centering
    \subfloat[]{\includegraphics[width=0.3\textwidth]{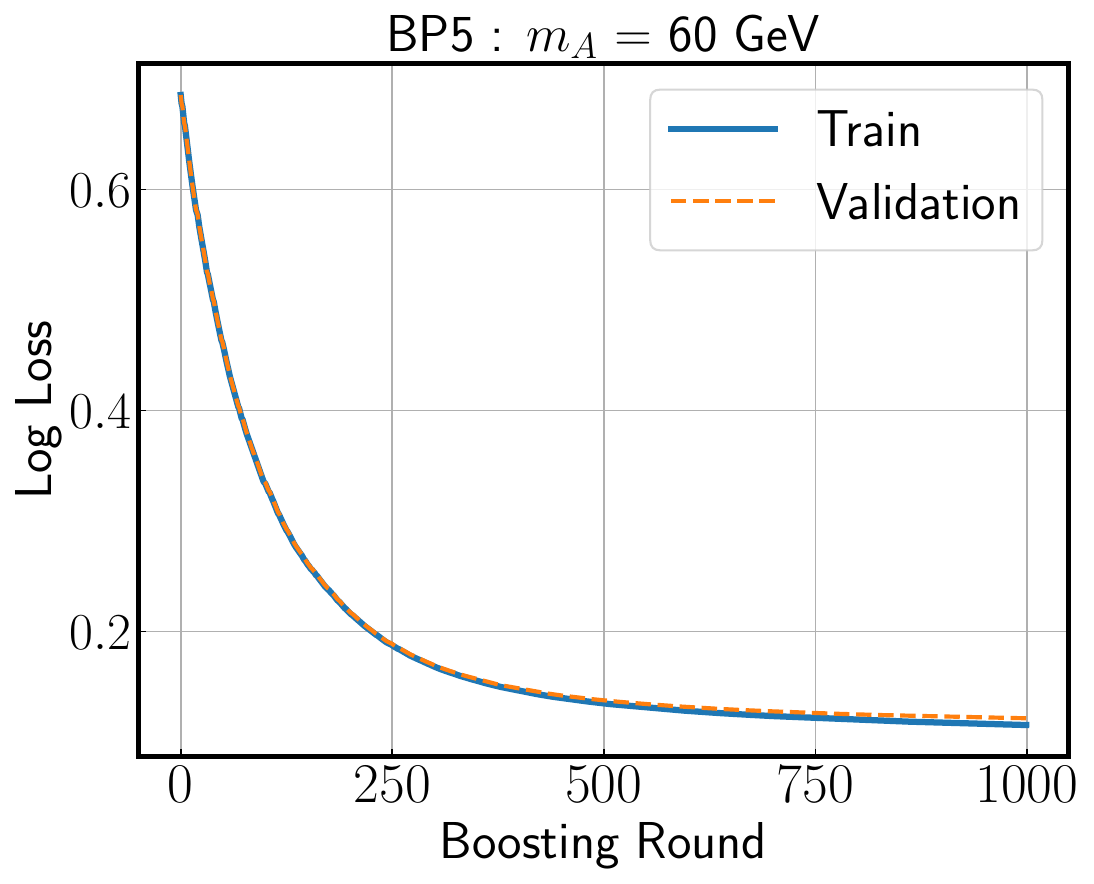}}\hfill
    \subfloat[]{\includegraphics[width=0.345\textwidth]{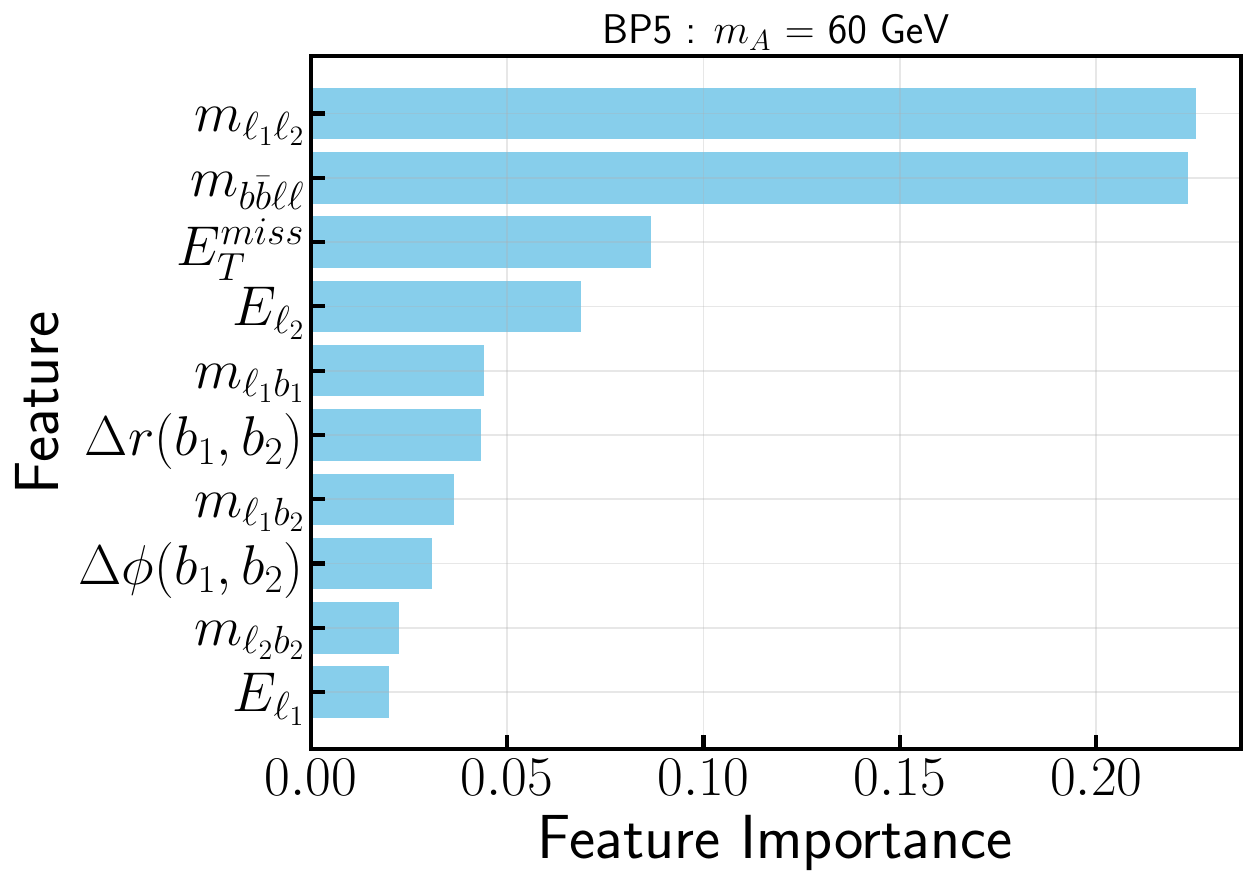}}\hfill
    \subfloat[]{\includegraphics[width=0.3\textwidth]{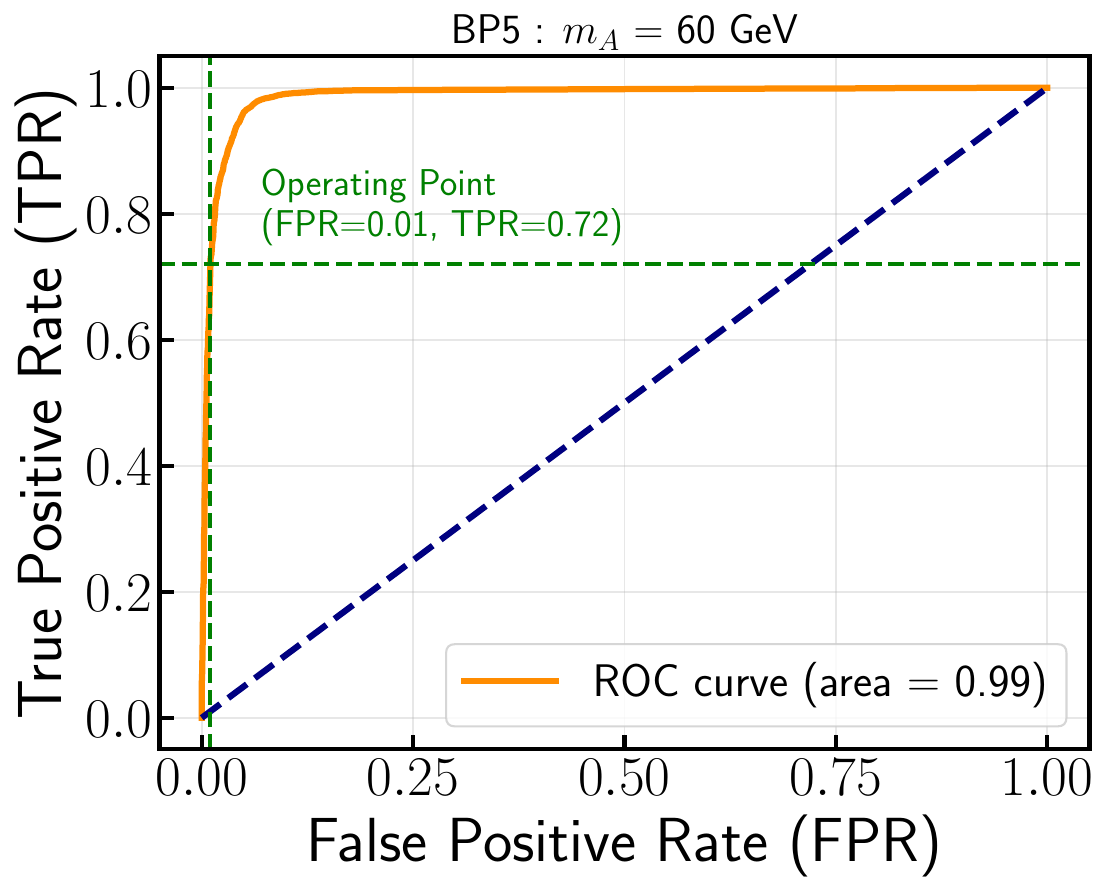}}

    \captionsetup{justification=raggedright,singlelinecheck=on}
    \caption{\it For BP5, $m_A = \text{60 GeV}$. The left plot shows the trained model learning curve on the training and validation datasets. The importance of the different input features during the training of BDT to discriminate the signal and background is shown in the middle plot.
    The right plot illustrates the Receiver Operating Characteristic (ROC curve) after training the BDT, with the area under the curve (AUC) of 0.99.}
    \label{fig:FeatureImpROC}
\end{figure}

The simulated dataset is split into 60\% for training, 25\% for test purposes, and another 15\% for validation during training. During the training of BDT, the discrimination of signals from a background heavily relies on selecting the input variables. So, after training, a feature importance ranking could be extracted from the trained model, which provides insight into which kinematic observables contribute the most to classification. The left plot of Figure~\ref{fig:FeatureImpROC} illustrates the learning curve of the BDT, demonstrating that the model is well-trained and does not exhibit overfitting. The middle plot of Figure~\ref{fig:FeatureImpROC} represents the importance of the variables BDT used to classify the signal and backgrounds. The three most significant variables turn out to be $m_{\ell \ell}$, $m_{b \bar{b} \ell \ell}$ and $E_T^{\rm miss}$. The right panel of Figure~\ref{fig:FeatureImpROC} shows the Receiver Operating Characteristic (ROC) curve, with an area under the curve (AUC) to be 0.99. This indicates the classification power of the trained BDT model is high, i.e., one can have a strong background rejection without losing much of the signal efficiency.

\begin{figure}[h!]
    \centering
    \subfloat[]{\includegraphics[width=0.475\textwidth]{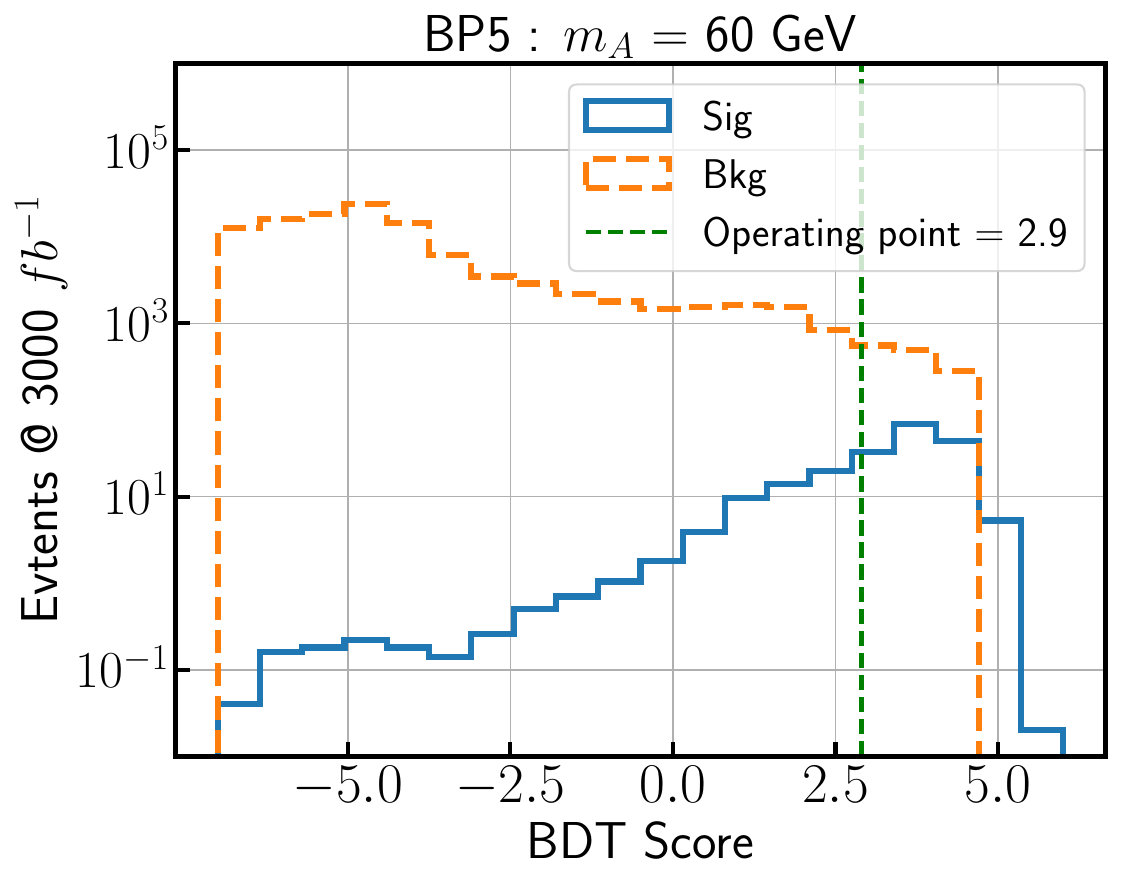}}\hfill
    \subfloat[]{\includegraphics[width=0.515\textwidth]{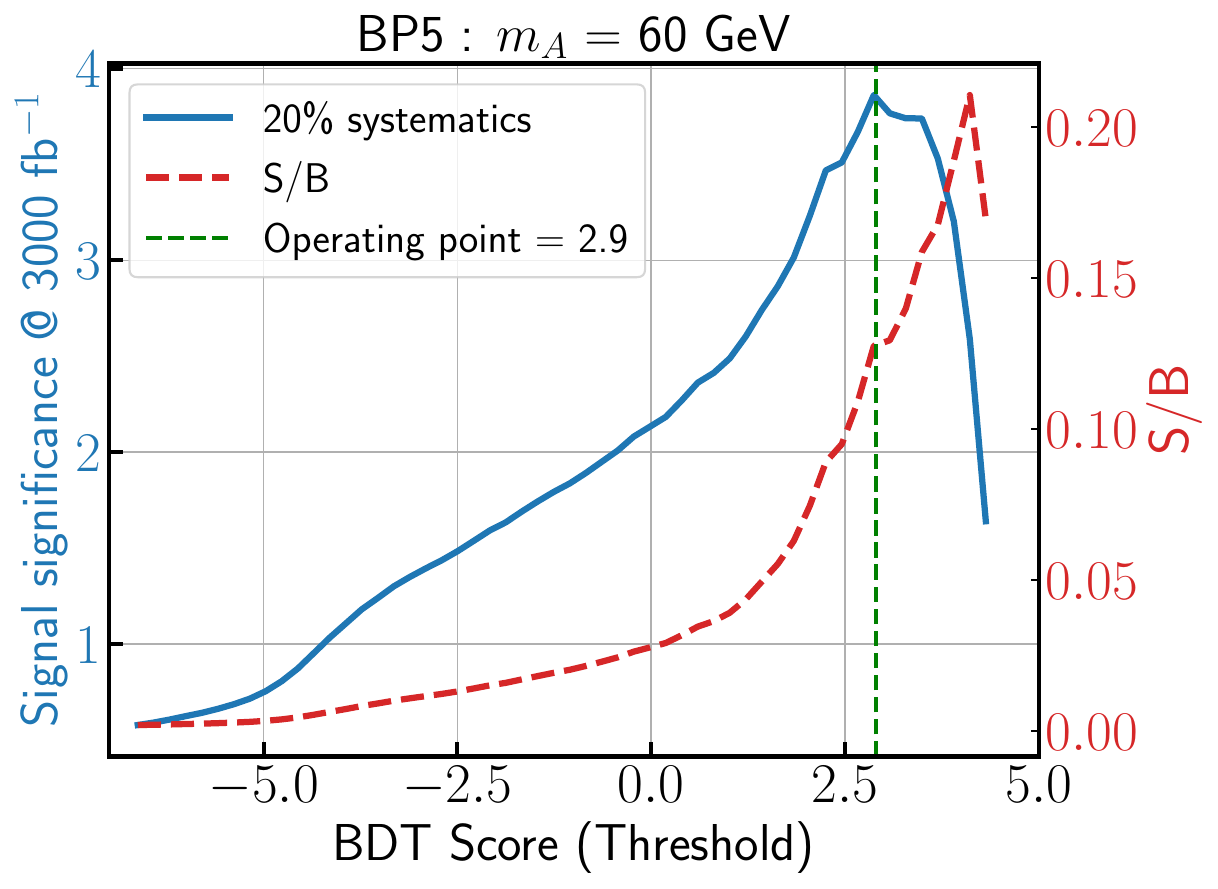}}
    \captionsetup{justification=raggedright,singlelinecheck=on}
    \caption{\it For BP5: $m_A = \text{60 GeV}$, the right panel represents the distribution of the number of signals and backgrounds at 3000 fb$^{-1}$ luminosity with BDT Score. The left panel shows the signal significance taking $20\%$ systematic uncertainty and the signal over background ratio at 3000 fb$^{-1}$ luminosity with BDT Score. The green vertical line in both plots indicates the operating point of BP5, which is set at 2.9.}
    \label{fig:SigTag}
\end{figure}

\begin{table}[h!]\renewcommand\arraystretch{1.35}
\centering
\begin{tabular}{|c|>{\centering\arraybackslash}p{1.25cm}|
                   >{\centering\arraybackslash}p{1.25cm}|
                   >{\centering\arraybackslash}p{1.25cm}|
                   >{\centering\arraybackslash}p{1.25cm}|
                   >{\centering\arraybackslash}p{1.25cm}|}
\hline
\textbf{} & \multicolumn{5}{c|}{\textbf{Significance ($\mathcal{S}$) at 20\% systematics}} \\
\cline{2-6}
\textbf{} & \textbf{BP1} & \textbf{BP2} & \textbf{BP3} & \textbf{BP4} & \textbf{BP5} \\
\hline
\textbf{Signal Significance} & 6.8 & 34 & 26 & 12 & 3.8 \\
\hline
\multicolumn{6}{|c|}{\textbf{Required Luminosity in $fb^{-1}$ for 5$\sigma$ significance}} \\
\hline
\textbf{} & 1622 & 65 & 111 & 520 & 5194 \\
\hline
\end{tabular}
\captionsetup{justification=raggedright,singlelinecheck=on}
\caption{\it This table summarizes the signal significance using pre-trained BDT models individually for each benchmark. The last row is the scaled luminosity for each benchmark point at the discovery level (5$\sigma$).}
\label{tab:signiBDT}
\end{table}
Using the trained BDT model, we evaluate the classification performance of the signal and background events by testing them separately. BDT predicts a score (BDT score) for each event (both signal and background). The left panel of Figure~\ref{fig:SigTag} shows the BDT scores distribution for the signal and backgrounds, indicated with blue and orange lines, respectively. As expected, the signal events predominantly clustered at the higher BDT scores, while the backgrounds accumulated at the lower. Then, we evaluate the signal significance by varying the BDT score as a threshold at an integrated luminosity of 3000 fb$^{-1}$, considering 20\% systematic uncertainties. The right plot of Figure~\ref{fig:SigTag} illustrates the signal significance variation with the BDT score threshold choice for the BP5, $m_A = \text{60 GeV}$ indicated by the blue line, while the signal-to-background ratio is plotted in the red line. In the cut-based approach, the signal significance is initially around 1.6$\sigma$, taking 20\% uncertainty, but this BDT gives a better result with nearly 3.8$\sigma$ deviation, if we choose the threshold of the BDT Score around 2.9. In Table~\,\ref{tab:signiBDT}, we tabulate signal significance given by BDT models, trained and tested individually for each benchmark point at 20\% systematic uncertainty. We then point out the luminosity that is required to reach the discovery level (5$\sigma$). For BP2 and BP3, the required luminosity is quite low, 65 and 111 fb$^{-1}$, respectively, and has already been achieved in the LHC. Therefore, offline analysis of the already existing data can enable one to probe the corresponding regions of flipped 2HDM.

\section{Summary and Conclusion} \label{sec:summ}
We have explored the detectability of the light neutral pseudoscalar in flipped 2HDM at the LHC, which can lie in the mass range of 20-60 GeV, consistently 
with all current theoretical and experimental limits.  On scanning the regions
of the parameter space, which are consistent with these constraints, we identify a
set of benchmark points for our analysis. Since the pseudoscalar
 in such a scenario mostly decays dominantly in the channel $A \rightarrow b{\bar{b}}$, 
 a channel appropriate for LHC search turns out to be $pp \rightarrow h \rightarrow b{\bar{b}} \ell^+ \ell^-$, 
 with the opposite-sign same-flavour leptons coming from a real or virtual $Z$-boson. Furthermore, the strain on the energy budget of the lepton-pair put by the hardness 
requirement of the $b$-pair, showing an $m_A$-peak, mostly requires the leptons to come
from a virtual $Z$.

We have performed a cut-based analysis first, identifying suitable event selection criteria
that succeed in suppressing the backgrounds rather effectively. Our analysis has demonstrated 
that $m_A \simeq 20$ GeV has a somewhat poor signal significance in such an analysis,
mostly because of the softness of $b$-jets. For  $m_A \gsim 40$ GeV,
on the other hand, kinematic suppression tends to reduce the significance. Nonetheless,
one expects signals across the aforementioned range, from the marginal level to the discovery 
level at the high-luminosity LHC.  Our analysis has identified 
$m_A \approx 30$ GeV as the most optimistic from the signal point of view.

In the second part of our analysis, we have shown that a BDT algorithm improves our results 
significantly for all benchmark points. We have pointed out that one could reach the discovery level for certain benchmarks at Run II itself. This has motivated us to recommend dedicated searches in this channel using the existing and
impending data, while the remaining parts of the light pseudoscalar parameter space 
in flipped 2HDM retain their promise of discovery when the luminosity reaches the $ab^{-1}$ level

\section*{Acknowledgements}
The authors are thankful to the High Performance Computing (HPC) Cluster (Kepler) facility provided by the Department of Physical Sciences at IISER Kolkata. S.~S.~thanks to the Council of Scientific and Industrial Research (CSIR) for funding this project. We thank Rituparna Ghosh and Anirban Kundu for useful discussions. B.M., S.S., and R.S. acknowledge the hospitality of the Indian Association for the Cultivation of Science, where part of the project has been discussed and a portion of this paper was written.

\bibliographystyle{JHEP}
\bibliography{refs}
\end{document}